\documentclass[showpacs,showkeys,preprintnumbers,plb,amsmath,amssymb,floats,floatfix]{revtex4}
\pdfoutput=1
\usepackage{amsmath,amssymb,bm,natbib}
\usepackage[paperwidth=210mm, paperheight=297mm,centering,hmargin=3cm,vmargin=3.5cm]{geometry}
\usepackage{epstopdf}
\usepackage{hyperref}
\usepackage{color}
\usepackage{booktabs}
\usepackage{graphicx}
\usepackage{multirow}
\usepackage{slashed}
\usepackage{comment}
\usepackage{float}
\makeatletter

\newcommand{\Rmnum}[1]{\expandafter\@slowromancap\romannumeral #1@}

\makeatother
\begin{document}

\def\bib{\bibitem}
\def\be{\begin{equation}}
\def\ee{\end{equation}}
\def\beq{\begin{equation}}
\def\eeq{\end{equation}}
\def\beqar{\begin{eqnarray}}
\def\eeqar{\end{eqnarray}}
\def\barr{\begin{array}}
\def\earr{\end{array}}
\def\dis{\displaystyle}
\def\lsim{\:\raisebox{-0.5ex}{$\stackrel{\textstyle<}{\sim}$}\:}
\def\gsim{\:\raisebox{-0.5ex}{$\stackrel{\textstyle>}{\sim}$}\:}
\def\tilh{\tilde{h}}
\def\and{\qquad {\rm and } \qquad}
\def\vev{\small \em {\it v.e.v. }}
\def\p{\partial}
\def\ga{\gamma^\mu}
\def\slp{p \hspace{-1ex}/}
\def\sleps{ \epsilon \hspace{-1ex}/}
\def\slk{k \hspace{-1ex}/}
\def\slq{q \hspace{-1ex}/\:}
\def\prl#1{Phys. Rev. Lett. {\bf #1}}
\def\prd#1{Phys. Rev. {\bf D#1}}
\def\plb#1{Phys. Lett. {\bf B#1}}
\def\npb#1{Nucl. Phys. {\bf B#1}}
\def\mpl#1{Mod. Phys. Lett. {\bf A#1}}
\def\ijmp#1{Int. J. Mod. Phys. {\bf A#1}}
\def\zp#1{Z. Phys. {\bf C#1}}
\def\etal{ {\it et al.} }
\def\ie{ {\it i.e.} }
\def\eg{ {\it e.g.} }

\title{\bf \large Probing $WW\gamma$ coupling through $e^- \gamma \rightarrow \nu_e W^-$ at ILC}

\author{\bf Satendra Kumar} \email{satendra@iitg.ernet.in} 

\affiliation{
Department of Physics,
Indian Institute of Technology Guwahati,
Guwahati 781 039, India}

\author{\bf P. Poulose} \email{poulose@iitg.ernet.in}

\affiliation{
Department of Physics,
Indian Institute of Technology Guwahati,
Guwahati 781 039, India}

\begin{abstract}
The anomalous $WW\gamma$ coupling is probed through $e\gamma\rightarrow \nu W$ at the ILC. With a spectacular single lepton final state, this process is well suited to study the above coupling.
Cross section measurements can probe $\delta \kappa_\gamma$ to about $\pm 0.004$ for a luminosity of 100 fb$^{-1}$ at $500$ GeV center-of-mass energy with unpolarized electron beam. The limits derivable on $\lambda_\gamma$ from the total cross section are comparatively more relaxed. Exploiting the energy-angle double distribution of the secondary muons, kinematic regions sensitive to these couplings are identified. The derivable limit on $\lambda_\gamma < 0$ could be improved to a few per-mil,  focusing on such regions.  More importantly, the angular distributions at fixed energy values, and energy distribution at fixed angles present very interesting possibility of distinguishing the case of $\lambda_\gamma <0$ and $\lambda_\gamma \ge 0$. 
\end{abstract}


\maketitle

\newpage 
\section{Introduction}
With the recent discovery of the Higgs boson by the LHC \cite{atlas, Aad:2013wqa, Chatrchyan:2013lba, cms}, the Standard Model (SM) has reaffirmed itself as the theoretical explanation of elementary particle dynamics, including presenting a plausible picture of Electroweak Symmetry Breaking (EWSB) \cite{Dawson-EWSB, Djouadi:2005gi} through the Higgs mechanism. While this is so, and not withstanding the fact that SM has been extensively tested very successfully by many different experiments \cite{Buchmuller:2006zu}, it is widely believed that the SM is an effective theory, which needs to incorporate suitable modifications at large energies. Many expect that this large energy scale could be as small as a few TeV, which is being explored at the LHC.  Apart from direct measurements of the properties of the Higgs boson, like its interaction couplings with itself, as well as with gauge bosons and fermions, signature of EWSB could be probed by understanding the structure and values of gauge-boson self interactions. This is so, because the longitudinal degrees of freedom of the gauge bosons arise from the Higgs sector. The relation between the Higgs-gauge boson couplings and the trilinear gauge couplings are presented and studied in Ref.~\cite{Corbett:2013pja,Gonzalez-Fraile:2014cya}. The discovery of Higgs boson of about $126$ GeV mass, while establishing the Higgs mechanism as the method of EWSB, has opened an era of Higgs precision studies. Being a hadronic machine LHC has limitations to undertake precision studies. The International Linear Collider (ILC) proposed to collide high energy, high luminosity electrons and positrons has the mission of studying the SM at high precision and to look for signals beyond the standard model~\cite{ILC, Asner:2013psa}. Such a machine is well suited to an in-depth analysis of the gauge boson interactions\, within and beyond the SM. A large number of physics studies establish the fact that the potential of ILC is further enhanced by considering high energy photon-photon ($\gamma \gamma$) collisions as well as electron-photon ($e\gamma$) collisions, apart from the electron-positron ($e^+e^-$) collisions. 

It is obvious that, the $e\gamma$ and $\gamma\gamma$ colliders are better suited to study the photon couplings with other gauge bosons like the $\gamma WW$, $\gamma\gamma WW$, $\gamma ZZ$ and $\gamma\gamma Z$ \cite{couplings}. In this article we will focus on $e\gamma\rightarrow \nu W$ with $W\rightarrow l\bar\nu$. Signature of such an event is a single lepton with large missing energy. This process is sensitive to new physics effects including anomalous   $\gamma WW$ \cite{pheno:egamWnu, pheno:Godfrey}, composite fermion models \cite{pheno:Gregores}, etc. In the case of anomalous $\gamma WW$ couplings, this process has the advantage over $e^+e^-\rightarrow W^+W^-$ \cite{ww}, which is sensitive to both $\gamma WW$ and $WWZ$ couplings. Again, $\gamma\gamma\rightarrow W^+W^-$ \cite{gam-gam2ww, Gounaris:1995mc} is sensitive to $\gamma WW$, $\gamma\gamma WW$ and $\gamma\gamma Z$ couplings, revealing the edge of $e\gamma$ collider to study $\gamma WW$. In most of the previous studies, observables at the production level of the single $W$ is investigated, with the exception of Ref. \cite{pheno:Godfrey}, where the authors consider angular spectrum of the secondary leptons, also including the effect of off-shell $W$. Our main aim of this work is to study the possibility to exploit the secondary lepton spectrum including the energy and angular distributions to probe relevant new physics signals arising in anomalous $\gamma WW$ coupling.
 
In many models beyond the SM the quartic and triple-gauge boson couplings including $\gamma WW$are altered from their SM values. In a model independent approach, an effective Lagrangian with terms additional to the SM Lagrangian is considered in phenomenological and experimental studies \cite{Hagiwara}.  Relevant to the process considered here, the effective $\gamma WW$ vertex is commonly parametrized in terms of $\delta \kappa_\gamma$ and $\lambda_\gamma$, in the absence of CP violation (with vanishing SM values).  LEP constraints in single-parameter analysis (taking one parameter at a time, keeping the others at their SM values) gives bounds of $-0.105<\delta \kappa_\gamma<+0.069$, $-0.099<\delta \kappa_\gamma<+0.066$ and $-0.059<\lambda_\gamma<+0.026$, $-0.059<\lambda_\gamma<+0.017$ and two-parameter analysis limits their values to $-0.072<\delta \kappa_\gamma<+0.127$ and $-0.068<\lambda_\gamma<+0.023$ \cite{LEPconstraints, Schael:2013ita} at 95\% C.L. Tevatron constraints from $W\gamma$ process are not contaminated by other couplings, but are more relaxed compared the LEP constraints to give $-0.51<\delta \kappa_\gamma<+0.51$ and $-0.12<\lambda_\gamma<+0.13$ \cite{CDFconstraints}. The CMS constraints \cite{Chatrchyan:2011rr} from $W\gamma$ process are also more relaxed compared the LEP constraints.

Phenomenology of anomalous $\gamma WW$ coupling in the context of LHC as well as ILC has been carried out in a number of recent publications \cite{pheno:egamWnu, pheno:Godfrey, LHCstudies, ILCstudies, e-gamma collision}.
 In particular, \cite{pheno:Godfrey} has analyzed single $W$ production with its leptonic decay to probe the effect of anomalous couplings in $e\gamma$ collision. In this work we exploit the full potential of the secondary lepton spectrum to study the effect of $\gamma WW$ coupling in $e\gamma \rightarrow \nu W \rightarrow \nu (l\bar \nu)$.
 
In the next section we provide some details of the process and the observables used. In section 3 we present our numerical results, and finally summarize the study and present our conclusions in the last section.

\section{Analysis and discussion}
\begin{figure}[H]
\begin{center}
\begin{tabular}{c c}
\hspace{-18mm}
\includegraphics[angle=0,width=80mm]{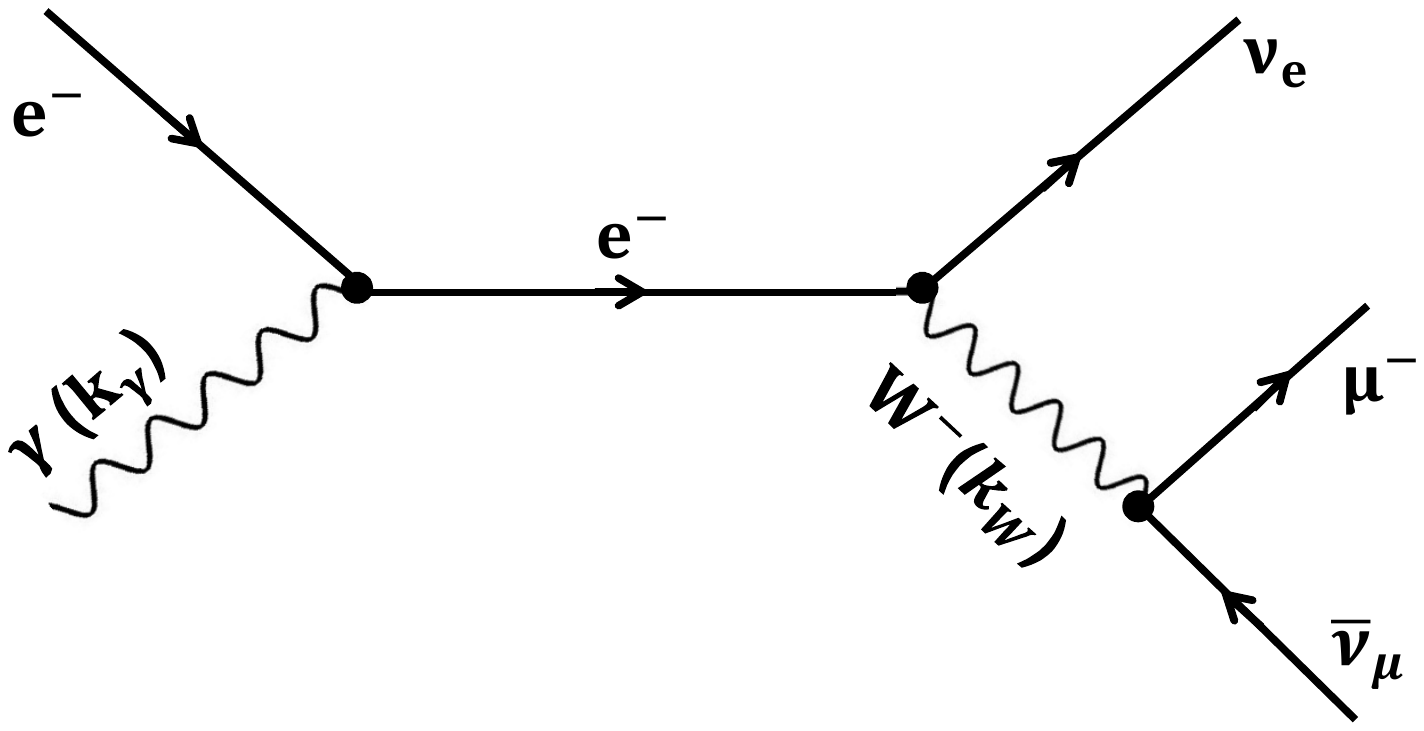} &
\hspace{-53mm}
\includegraphics[angle=0,width=80mm]{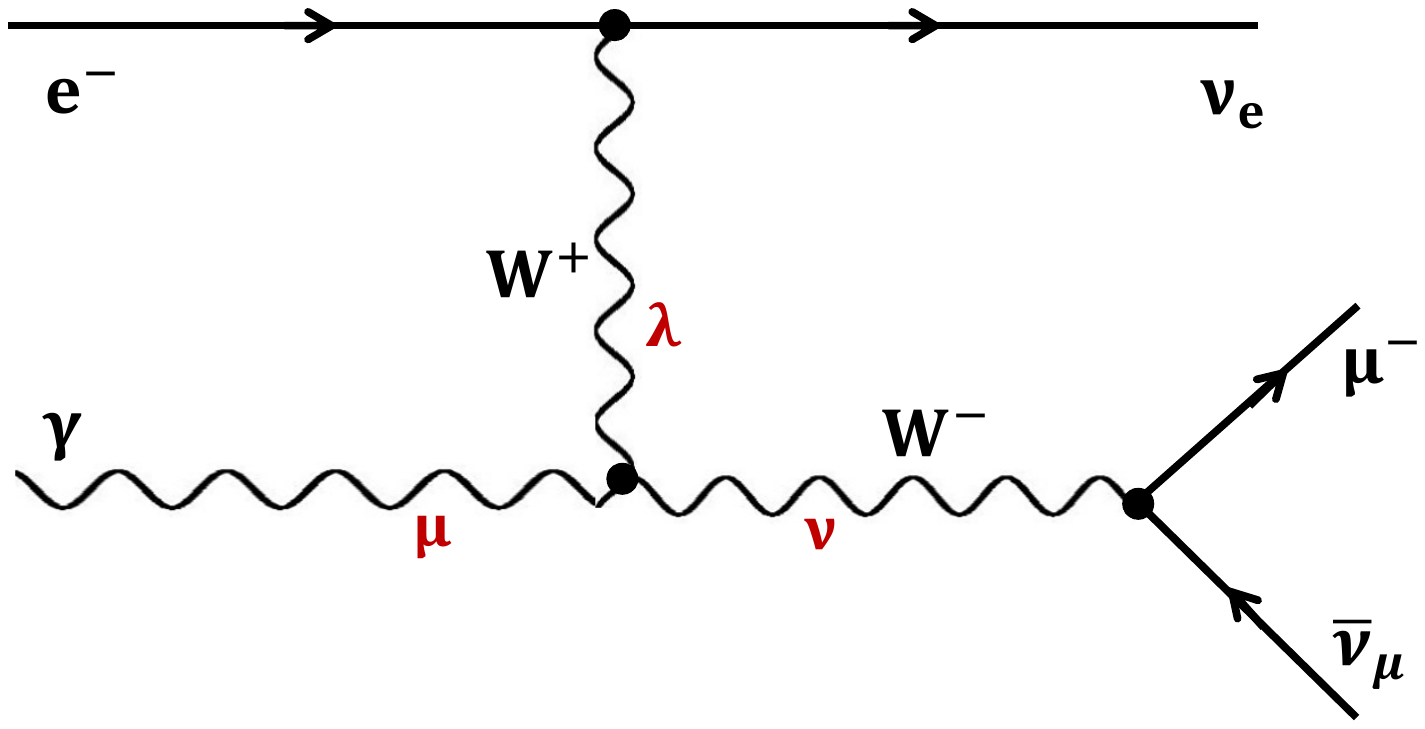}\\[-35mm]
\hspace{50mm}
\includegraphics[angle=0,width=80mm]{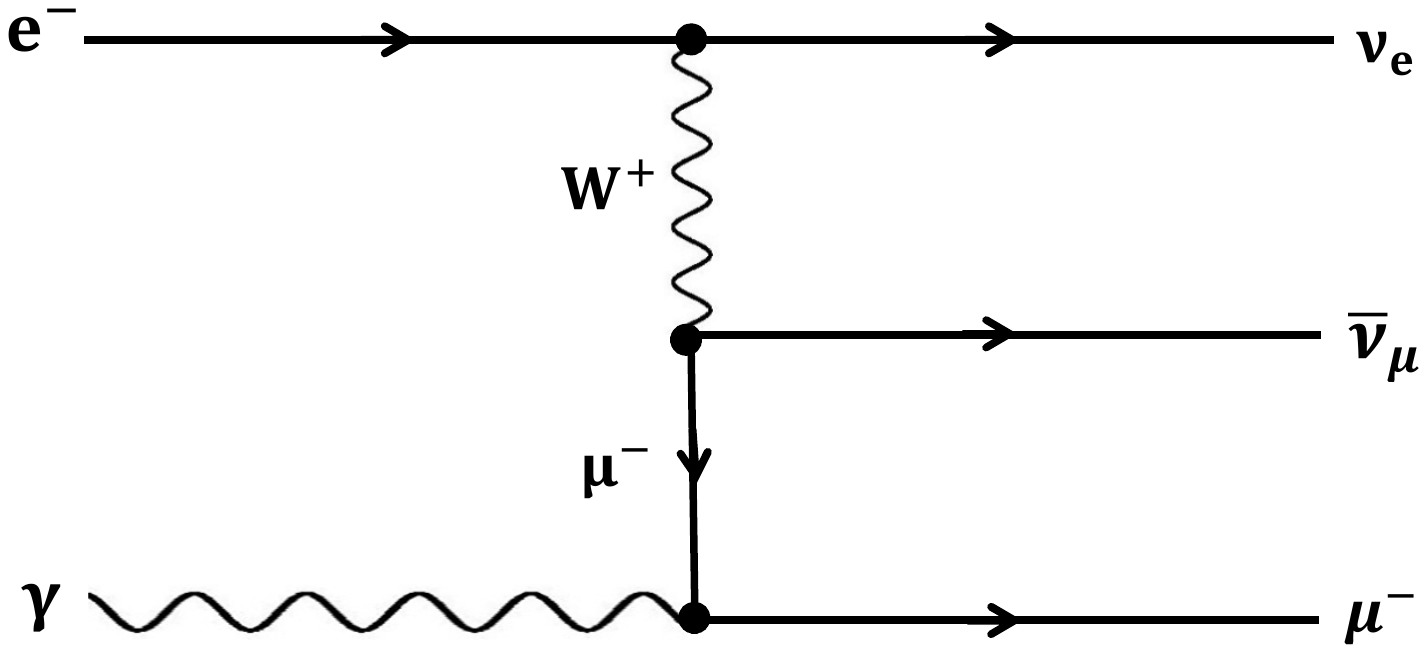}
\end{tabular}
\end{center}
\vspace{-36mm}
\caption{Feynman diagrams for the process $e\gamma\rightarrow \nu W\rightarrow \nu (\mu\bar \nu)$. First diagram shows the momenta used and second diagram shows the indices used in the vertex in Eq.~\ref{eqn:Vertex}}
\label{fig:Feyn}
\end{figure}

Considering a real on-shell photon, the most general CP-conserving $\gamma WW$ coupling within a  Lorentz invariant Lagrangian can be written in the following form \cite{Hagiwara}.
\begin{eqnarray}
{\cal L}_{\gamma WW}=-ie~\Big[W^\dagger_{\mu\nu}W^\mu A^\nu-W^\dagger_\mu A_\nu W^{\mu\nu} + (1+\delta \kappa_\gamma)~W^\dagger_\mu W_\nu F^{\mu\nu} + \frac{\lambda_\gamma}{m_W^2}~W^\dagger_{\lambda\mu}W^\mu_{\nu}~F^{\lambda\nu}\Big]
\label{eqn:Leff}
\end{eqnarray} 
 In the SM, the gauge structure $SU(2)_L\times U(1)_Y$ dictates the $\gamma WW$ couplings, with vanishing $\delta \kappa_\gamma$ and $\lambda_\gamma$ at tree level.  Therefore, precise measurements of these couplings will test the gauge sector of the electroweak interactions. Fig.~\ref{fig:Feyn} shows the Feynman diagrams for the process along with the momentum labels. 
 
With the effective Lagrangian given by Eq.~\ref{eqn:Leff}, the $\gamma WW$ vertex for the process under study (Fig.~\ref{fig:Feyn} $(b)$)  takes the form:
\begin{eqnarray}
\Gamma^{\mu\nu\lambda}&=&ie~\Bigg[  
2k_W^\mu~g^{\nu\lambda}+2k_\gamma^\nu~g^{\mu\lambda}-(k_\gamma+k_W)^\lambda~g^{\mu\nu} +(\delta \kappa_\gamma-\lambda_\gamma)(k_\gamma^\nu~g^{\mu\lambda} -k_\gamma^\lambda~g^{\mu\nu}) \\\nonumber 
&&+ \frac{\lambda_\gamma}{m_W^2} (k_\gamma+k_W)^\lambda( k_W^\mu k_\gamma^\nu-(k_\gamma\cdot k_W)g^{\mu\nu})\Bigg]
\label{eqn:Vertex}
\end{eqnarray}

The effective Lagrangian in Eq.~\ref{eqn:Leff} should be considered as a low energy approximation of some fundamental theory, which is expected to emerge at some high energy scale, $\Lambda$. 
To control unitarity violation at high energies, we consider the anomalous couplings as form factors according to the following \cite{pheno:Godfrey}
\begin{equation}
A =A_0/\left[\left(1+\frac{|k_W^2|}{\Lambda^2}\right)
\left(1+\frac{|(k_\gamma-k_W)^2|}{\Lambda^2}\right) \right],
\label{eqn:HEcompletion}
\end{equation}
where $A_0\equiv \delta \kappa_\gamma,~\lambda_\gamma$. 

Contribution from the third diagram (Fig.\ref{fig:Feyn}) is found to be a few percent, and will be neglected. This allows us to  perform our computation in the Narrow-Width Approximation (NWA) in which the $W$-propagator is approximated to get 
\begin{equation}
\frac{1}{|k_W^2-m_W^2|^2}=\frac{\pi}{m_W\Gamma_W}~\delta(k_W^2-m_W^2),
\label{eqn:NWA}
\end{equation}
 where $m_W$ is the mass and $\Gamma_W$ is the width of the $W$ boson.
 
 \begin{figure}[H]
\begin{center}
\includegraphics[angle=0,width=17cm]{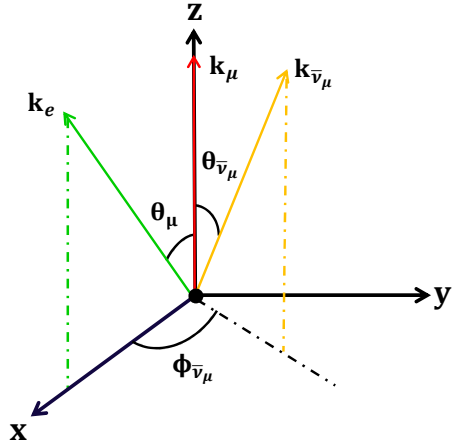}
\vspace{-6.4cm}
\caption{Reference frame defining different momenta and corresponding angles used in Eq.~\ref{eqn:ddist0}.}
\label{fig:RF}
\end{center}
\end{figure}

To perform phase space integrations we fix our reference frame as the center-of-mass frame (CMF) of the colliding electron and photon system. $z$-axis is taken along $\vec{k_\mu}$, which is the momentum of the outgoing lepton (considered as muon in the further discussion) as pictured in Fig.~\ref{fig:RF}. $y$-axis is defined as $\vec{k_\mu}\times \vec{k_e}$, where $\vec{k_e}$ is the momentum of the colliding electron. The $\bar\nu_\mu$ comes out at a polar angle $\theta_{\bar\nu_\mu}$ and azimuthal  angle $\phi_{\bar\nu_\mu}$. Energy-momentum conservation and the NWA (Eq.~\ref{eqn:NWA}) are used to get the differential cross-section
\begin{equation}
\frac{d\sigma}{dx_\mu~d\cos\theta_\mu~d\phi_{\bar\nu_\mu}}=\frac{x_\mu}{(2\pi)^3~128~m_W~\Gamma_W}|M_r|^2,
\label{eqn:ddist0}
\end{equation}
where $x_\mu=\frac{2E_\mu}{\hat s}$, with $\sqrt{\hat s}$ the center-of-mass energy and $E_\mu$ is the energy of the muon, and $\cos\theta_\mu= \frac{\vec{k_\mu}\cdot \vec{k_e}}{|\vec{k_\mu}\cdot \vec{k_e}|}$. 
Here $M_r$ is the reduced amplitude given in terms of the invariant amplitude $M$ as,
\begin{equation}
M = \frac{1}{(k_W^2-m_W^2)}~M_r.
\end{equation}
$|M_r|^2$ is obtained using FORM computational package \cite{FORM}. After integrating the unobservable $\phi_{\bar\nu_\mu}$, we get the double distribution of energy and polar angle of the secondary muon in the center-of-mass frame of the colliding particles with the electron momentum now taken along the redefined $z$ axis. Notice that the muon energy in the CMF is bounded by $\frac{m_W^2}{2\sqrt{\hat s}} \le  E_\mu \le \frac{\sqrt{\hat s}}{2}$.

To obtain the distribution in the lab frame, we need to boost the above differential cross-section appropriately. For an electron beam energy of $E_e$ and photon energy of $\omega_\gamma=x~E_e$, we have the following relation between the variables in the CMF and the laboratory frame.
\begin{eqnarray}
\hat s &= &x~4E_e^2\nonumber\\
E_\mu&=&E_\mu^{lab}~\gamma~\left(1-\beta~\cos\theta^{lab}_\mu\right) \nonumber \\
\cos\theta_\mu&=&\frac{\cos\theta^{lab}_\mu-\beta}{1-\beta~\cos\theta^{lab}_\mu},
\label{eqn:lab2cmf}
\end{eqnarray}
where $\beta$ is the speed of the CMF compared to the lab frame, and $\gamma=\frac{1}{\sqrt{1-\beta^2}}$. Notice that the limits of $E_\mu^{lab}$ integration depends on $\cos\theta^{lab}_\mu$, keeping it within the bound 
\begin{equation}
\frac{m_W^2}{4\sqrt{x}E_e} \frac{1}{\gamma \left(
1-\beta \cos\theta^{lab}_\mu\right)}\le  E_\mu^{lab} \le\frac{ \sqrt{x}E_e}{\gamma \left(
1-\beta \cos\theta^{lab}_\mu\right)}.
\label{eqn:boundsElab}
\end{equation}
It is well known that the hard photons produced by compton scattering are polarized. The polarization depends on the initial electron and laser beam polarizations. Besides, the final photon distribution will depend on the initial beam energies. Considering these, the parton level cross section need to be folded with the appropriate luminosity function, ${\cal L}_{e\gamma}(x)$, an expression for which is provided in the Appendix, so that the total cross-section in the lab frame is given by
 \begin{equation}
\sigma = \int \frac{d{\cal L}_{\gamma/e}(x)}{dx}~\sigma(\hat s)~dx.
\label{eqn:lumfold}
\end{equation}

We suitably use Eqs.~\ref{eqn:ddist0}, \ref{eqn:lab2cmf}, \ref{eqn:lumfold} to obtain the total cross-section, the muon angular distribution $\frac{d\sigma}{d\cos\theta^{lab}_\mu}$, the scaled muon energy distribution $\frac{d\sigma}{dx^{lab}_\mu}$, and the energy-angle distribution of the muons \( \frac{d\sigma}{dx^{lab}_\mu~d\cos\theta^{lab}_\mu}\) in the lab frame.
Integrations over the photon distribution variable $x$, the scaled muon energy $x^{lab}_\mu$ and the muon angular variable $\cos\theta^{lab}_\mu$ in appropriate cases are performed numerically using the {\tt Cuhre} routines under the {\tt CUBA} package \cite{CUBA}. From now on we will drop the superscript {\small \em $lab$} from the variables.

Phenomenological analysis of $e\gamma\rightarrow \nu W$ is considered by some authors in the past \cite{pheno:egamWnu, pheno:Godfrey}. Most of the studies limit their analysis at the production level. Experimentally it is more useful to understand the effect on the final state particles arising from $W$ decay. This is all the more important in the case of leptonic decay, as it is not possible to reconstruct the $W$ in such case.  In ref.~\cite{pheno:Godfrey}, analysis including decay spectrum is presented, where detailed study of the secondary lepton angular distribution is considered along with other reconstructed observables concerning the $W$ production.  In the analysis presented here we demonstrate the usefulness of the combined energy-angle distribution of the secondary leptons in extracting information on the anomalous $\gamma WW$ couplings. While Ref.~\cite{pheno:Godfrey} takes into account the contribution due
to off-shell $W$ along with the on-shell production, our work is in the NWA assuming on-shell $W$ production. At the same time, as note above, the effect of off-shell contribution, mainly arising through the diagram with muon propagator, is negligible, while the our analysis with energy-angle double distributions, along with combinations of the angular and energy distributions gives a handle on disentangling the effect of $\delta \kappa_\gamma$ and $\lambda_\gamma$, to certain extend.

\section{Numerical Results}

For our numerical analysis we consider ILC of center-of-mass energies of $500$ GeV and $1000$ GeV, with the option of $e\gamma$ collision using backscattered laser photons, and with unpolarized electron beams. It may be noted that as the process being considered is a purely a weak interaction process, only left-handed electrons will take part. Therefore, the left-polarized electron beam would enhance the cross-section, without presenting any additional advantage. We shall present all our analysis for unpolarized beam, without losing any generality.

\begin{figure}[h]
\begin{center}
\includegraphics[angle=0,width=10cm]{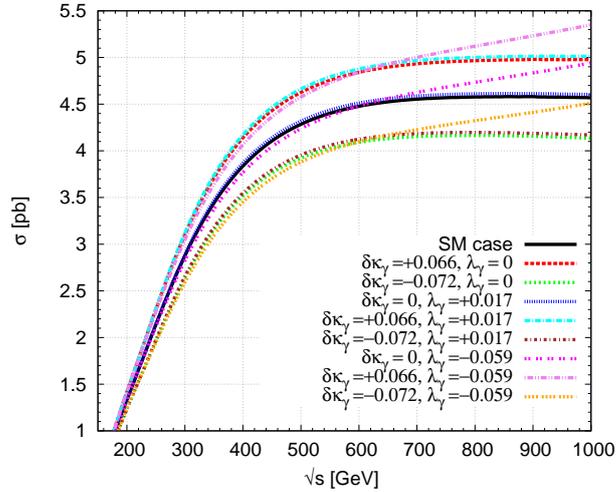}
\vspace{-8mm}
\caption{{ Total cross-section against $\sqrt{s}=2E_e$, where $E_e$ is the electron beam energy, for different anomalous couplings, along with the SM case.}}
\label{fig:sigvsroots1}
\end{center}
\end{figure}

\begin{figure}[h]
\begin{center}
\begin{tabular}{c c}
\hspace{-3mm}
\includegraphics[angle=0,width=75mm]{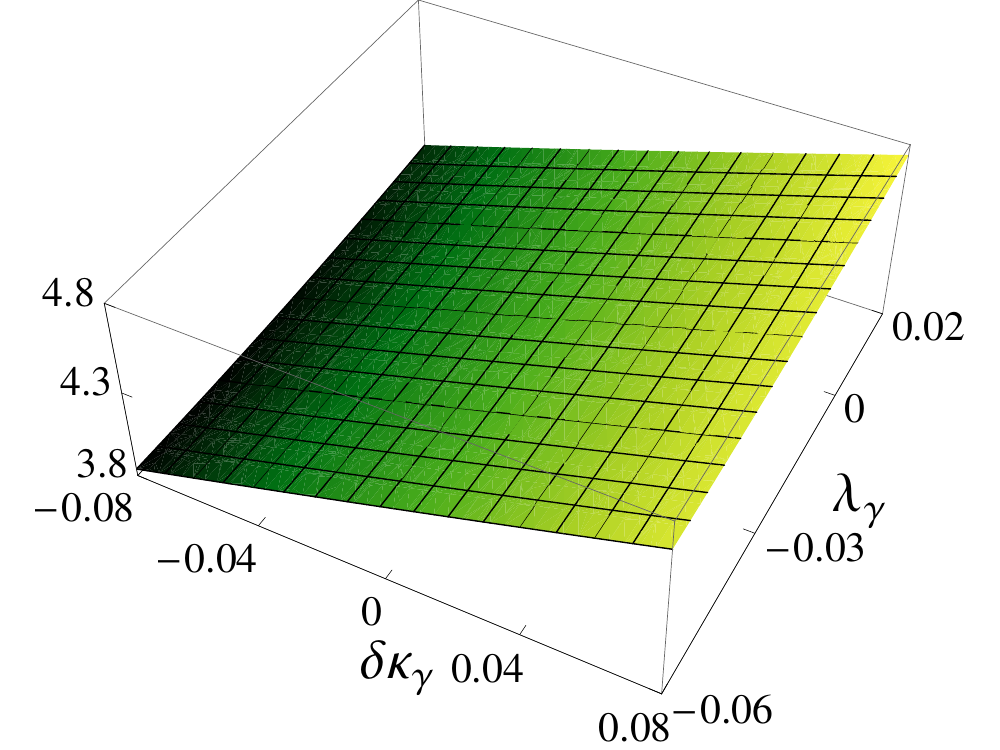} 
\hspace{-1mm}
\includegraphics[angle=0,width=75mm]{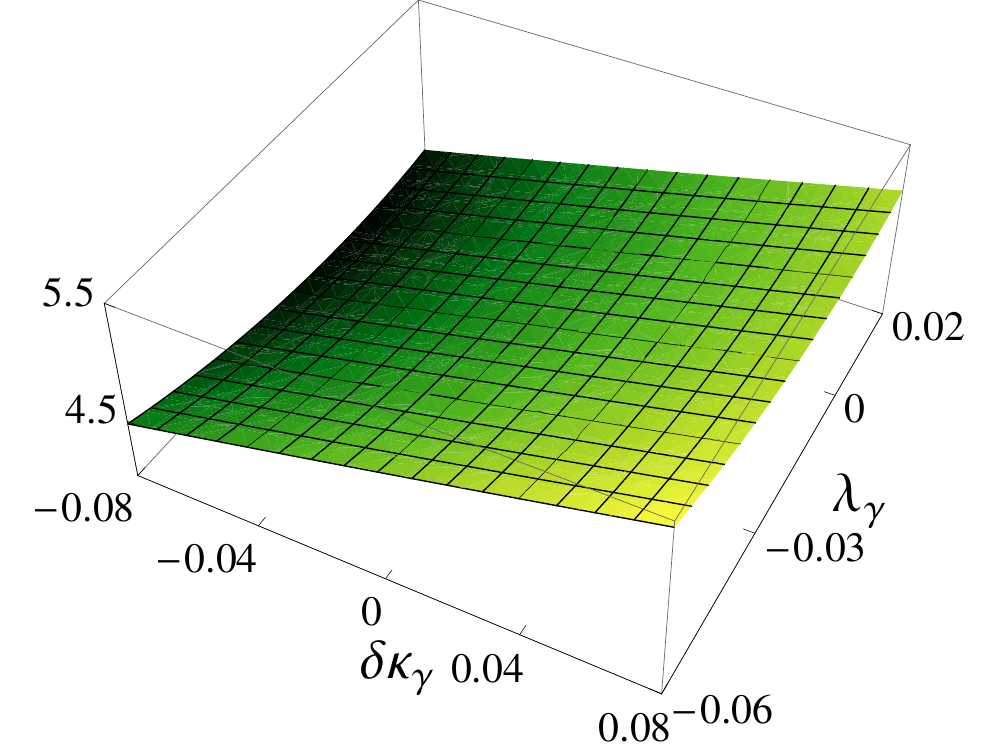}
\end{tabular}
\caption{{ The surface plots showing the total cross-section against $\delta\kappa_{\gamma}$ and $\lambda_{\gamma}$ for the center-of-mass energy ($\sqrt{s}$) of $~500$ GeV (left) and $1000$ GeV (right), where cross-section in $pb$ is along the vertical axis.}}
\label{fig:sigvsroots5}
\end{center}
\end{figure}

Firstly, we consider the total cross-section. Fig.~\ref{fig:sigvsroots1}, shows the total cross-section against the center-of-mass energy of the $e^-e^-$ system (denoted as $\sqrt{s}$), one of which Compton-scatters on the laser beam to produce the high energy photon beam.  The effect of $\delta \kappa_\gamma$ has an effect of about 7\% to 8\% almost all through the different center-of-mass energy values, while the effect of $\lambda_\gamma$ has a dependence on the $\sqrt{s}$. This is expected, as the former is the coefficient of a dimension-4 operator, while the latter is the coefficient of a dimension-6 operator, thus producing a momentum dependence in the corresponding coupling. Again, for the same order of magnitudes of the couplings, the cross section is more sensitive to $\delta \kappa_\gamma$ compared to $\lambda_\gamma$, as expected for higher dimensional operators. One may note that the sensitivity to the sign of $\lambda_\gamma$ ceases beyond the center-of-mass energies slightly beyond 500 GeV. This means that the effect of the interference term is overshadowed by the contribution purely coming from the dimension-6 anomalous term, which goes like the square of $\lambda_\gamma$.  
In order to be more quantitative, in Table~\ref{table:cross-section}, we have presented the cross-sections for the limiting values of the couplings for two selected center-of-mass energy values, $\sqrt{s}=500$ GeV and $1000$ GeV. Taking individually separately, the effect of $\lambda_\gamma$ is small, with a maximum of about 7\% for negative values at $1000$ GeV center-of-mass energy. It may be noted that the effect is sensitive to the sign of the parameter, indicating that the interference term is dominant, as expected. At $500$ GeV, one can ignore the effects of $\lambda_\gamma$ even when considering along with $\delta \kappa_\gamma$, whereas at $1000$ GeV, the presence of $\lambda_\gamma$ can either nullify or enhance considerably the effect of $\delta \kappa_\gamma$. Largest effect is seen when $\delta \kappa_\gamma$ is positive and $\lambda_\gamma$ is negative, with about 20\% deviation in the total cross-section at $1000$ GeV. 
In the surface plots shown in Fig.\ref{fig:sigvsroots5}, variation of the cross-section with $\delta\kappa_\gamma$ and $\lambda_\gamma$ at center-of-mass energy of $500$ GeV and $1000$ GeV are presented, which show monotonous dependences, almost a linear one, as expected from the fact that only the interference terms play role. We shall see below that, with such large cross-section, small deviations even at percent or sub-percent level will be able to probe the relevant parameters significantly.

\begin{center}
\begin{table}[h]
\begin{tabular}{|c|c|c|}
\hline
   \textbf{$\delta\kappa_\gamma, ~~~~~~~\lambda_\gamma$}  & \multicolumn{2}{c|} {\textbf{$~~~~\sigma$} [pb]  } \\\cline{2-3}
                  & $\sqrt{s}=500$  GeV &$\sqrt{s}=1000$  GeV  \\
\hline\hline
   SM case               & 4.284     & 4.568   \\\cline{2-3}            
\hline\hline
  +0.066,~~~~~~~~ 0      & 4.626     & 4.977  \\\cline{2-3}
\hline
  -0.072,~~~~~~~~~ 0     & 3.922     & 4.135  \\\cline{2-3}
\hline  
   ~~~~0,~~~~~~~+0.017   & 4.319     & 4.602  \\\cline{2-3}
\hline    
   ~~~~0,~~~~~~~~-0.059  & 4.238     & 4.942  \\\cline{2-3}
\hline 
  +0.066,~~~ +0.017      & 4.662     & 5.012  \\\cline{2-3}
\hline   
  +0.066,~~~~-0.059      & 4.576     & 5.350  \\\cline{2-3}
\hline   
  -0.072,~~~~~+0.017     & 3.955     & 4.169  \\\cline{2-3}
\hline
  -0.072,~~~~~-0.059     & 3.879     & 4.510  \\\cline{2-3}
\hline
\end{tabular}
\caption{{Total cross-section for different combinations of TGC parameters at $\sqrt{s}=500$ GeV and $1000$ GeV.}}
\label{table:cross-section}
\end{table}
\end{center}
 
\begin{figure}[h]
\begin{center}
\includegraphics[angle=0,width=10cm]{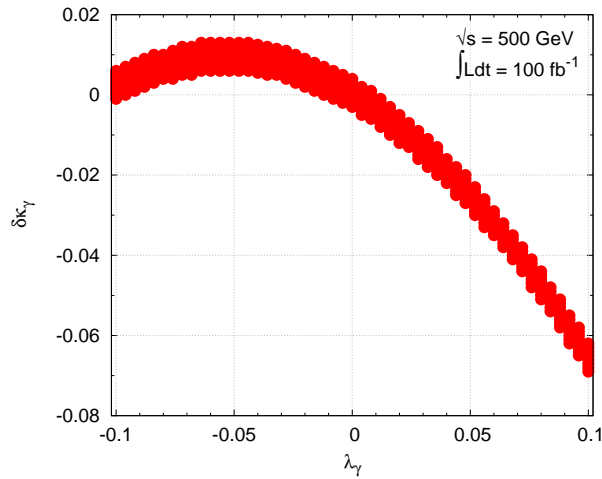}
\vspace{-8mm}
\caption{{The shaded region corresponds to values of $\delta\kappa_{\gamma}-\lambda_{\gamma}$ with the total cross-section within the $3\sigma$ limit, for an integrated luminosity of $100~fb^{-1}$ at a center-of-mass energy of $500$ GeV.}}
\label{fig:correlation}
\end{center}
\end{figure}

The two-parameter bound showing the allowed region in the $\delta \kappa_\gamma - \lambda_\gamma$ plane is presented in Fig.\ref{fig:correlation},  at center-of-mass energy of $500$ GeV with an integrated luminosity of $100~fb^{-1}$.This corresponds to number of event of about 430000. Thus, with the dependence described above, the reach on the parameters for these machine parameters is expected to be very good.
 We remind the reader that, when cross-section is considered as a function of anomalous coupling parameters $\delta\kappa_\gamma$ and $\lambda_\gamma$, the result is a second order polynomial in these two parameters. With this, the $3\sigma$ limit of the cross-section leads to an elliptic equation corresponding to the relation between these two parameters. This results in an elliptic band in the $\delta\kappa_\gamma-\lambda_\gamma$ plane respecting the $3\sigma$ limit of the cross-section. Very strong bound on $\delta \kappa_\gamma$  can be derived for any assumed value of $\lambda_\gamma$. On the other hand, for small values of $\delta\kappa_\gamma$ the sensitivity of cross-section on $\lambda_\gamma$ is comparatively smaller. At the same time, the relative sign of the parameters is preferred to be opposite to each other, as was also evident from the analysis above.

\begin{figure}
\begin{center}
\begin{tabular}{c c}
\hspace{-6mm}
\includegraphics[angle=0,width=88mm]{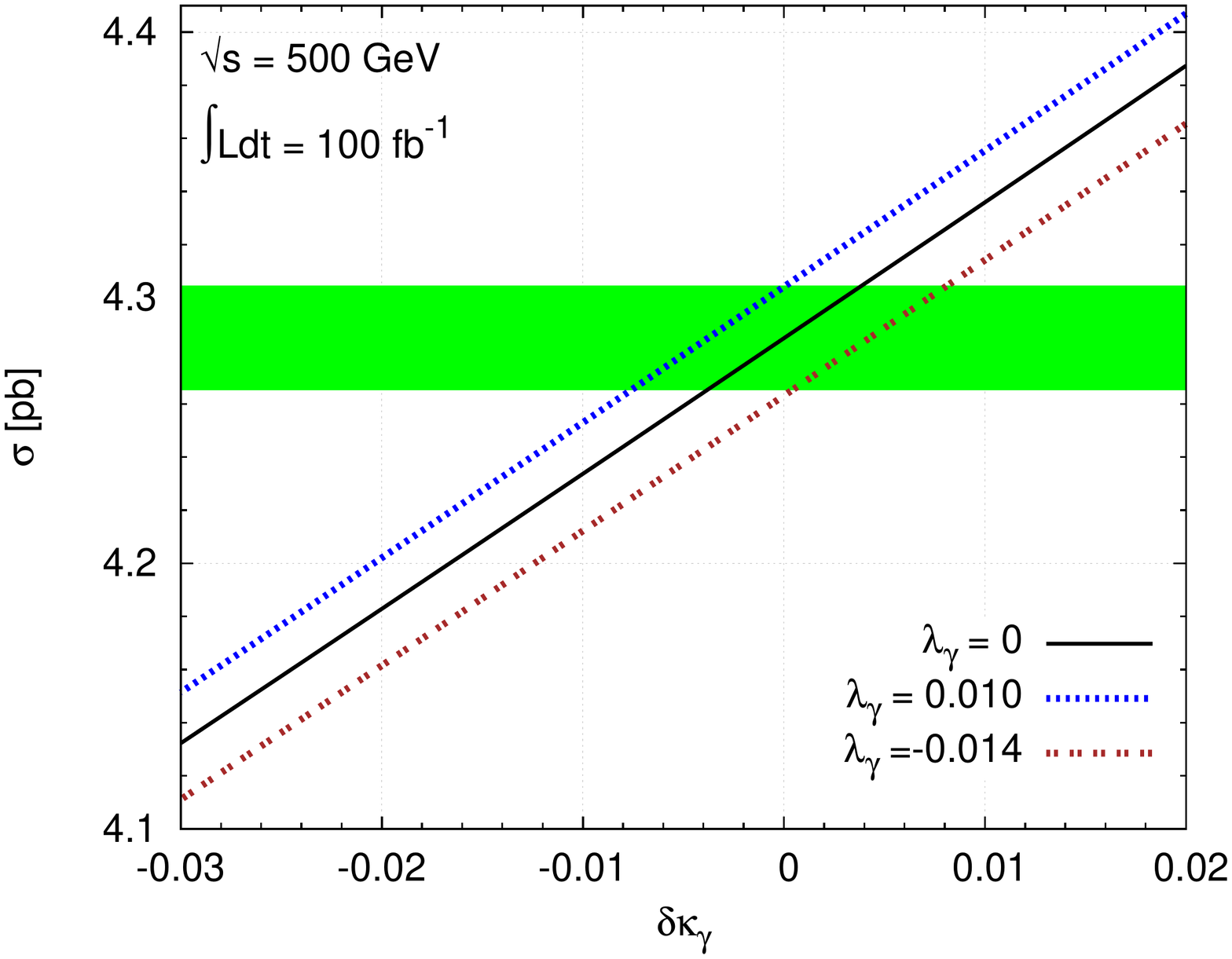} 
\hspace{-14mm}
\includegraphics[angle=0,width=88mm]{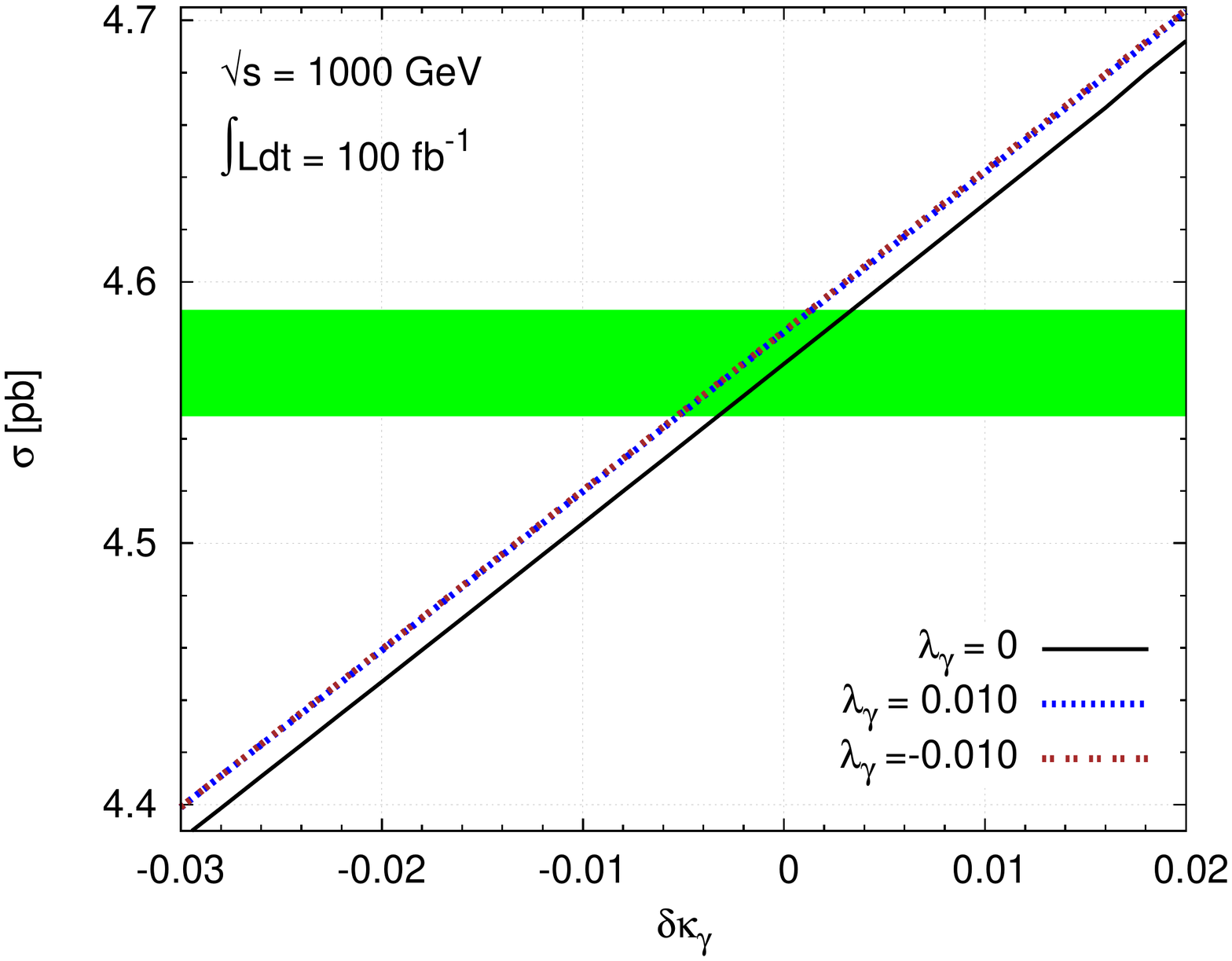}\\
\hspace{-6mm}
\includegraphics[angle=0,width=88mm]{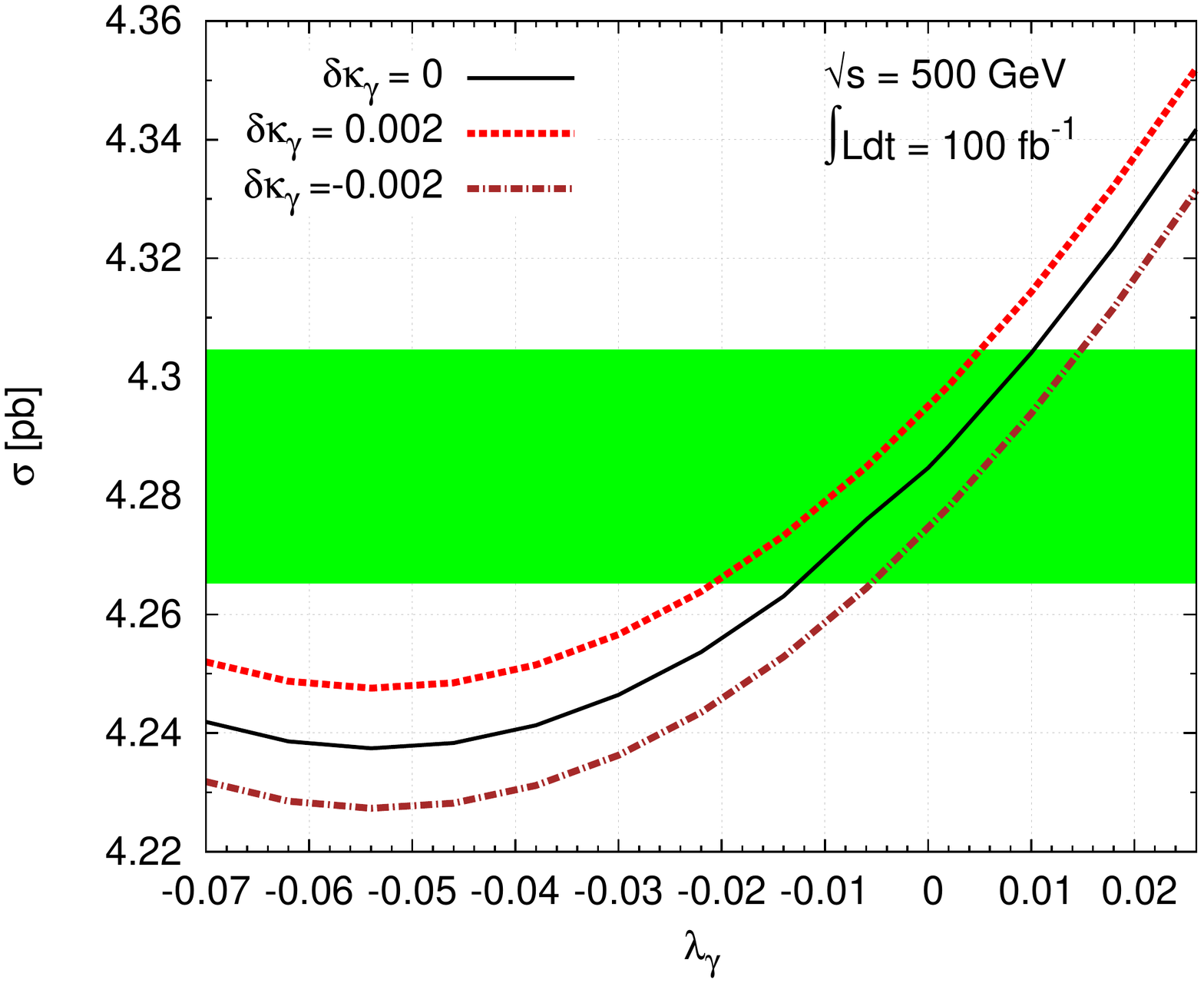}   
\hspace{-14mm}
\includegraphics[angle=0,width=88mm]{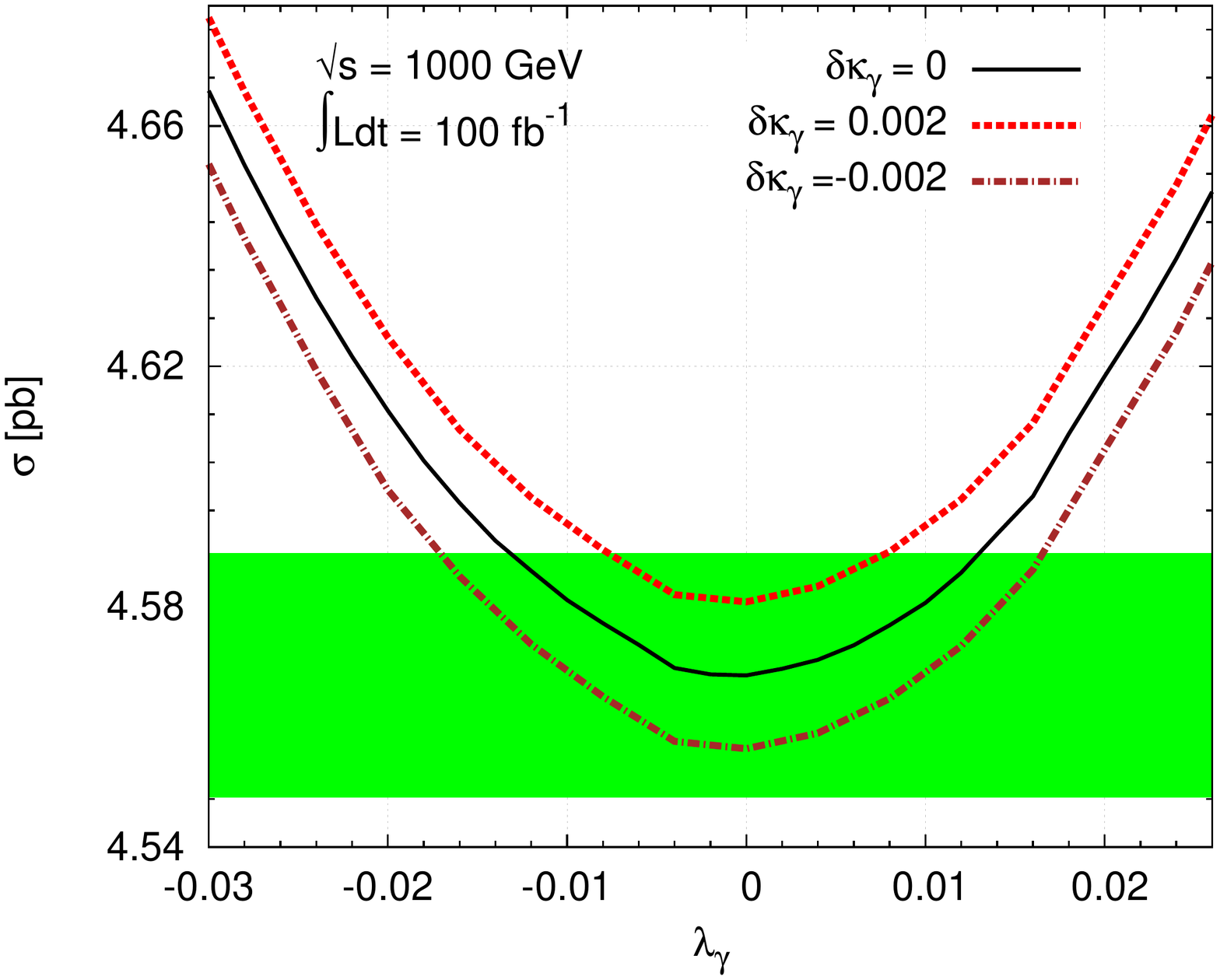} 
\vspace{-8mm}
\end{tabular}
\caption{{Cross-section against $\delta \kappa_\gamma$ (top row)  and $\lambda_{\gamma}$ (bottom row), when the other parameter assume typical values.  The center-of-mass energies considered are $\sqrt{s}=500$ GeV (left column) and $1000$ GeV (right column). The green band indicates the $3\sigma$ limit of the SM cross-section, with integrated luminosity of 100 fb$^{-1}$.}}
\label{fig:cs_dkp_lmd}
\end{center}
\end{figure}

Moving on to the single parameter, sensitivity of cross-section on one parameter, while the other one is fixed is presented in Fig. \ref{fig:cs_dkp_lmd}, for center-of-mass energies of $500$ GeV and $1000$ GeV.  The green band represents the $3\sigma$ region of the SM value of the cross-section, considering an integrated luminosity of $100~$fb$^{-1}$. The cross-section has a linear dependence on $\delta \kappa_\gamma$, showing that the contribution proportional to the quadratic term is negligible. The $3\sigma$ allowed range of $\delta \kappa_\gamma$ shifts with the value of $\lambda_\gamma$. Although, as Table~\ref{table:cross-section} suggests, the effect of $\lambda_\gamma$ is small compared to the large effect of $\delta \kappa_\gamma$ values,  the influence is significant for derivable limits as can be read from Fig.~\ref{fig:cs_dkp_lmd}. For example, at $500$ GeV center-of-mass energy, assuming $\lambda_\gamma=0$, one obtain a $3\sigma$ limit of 
$-0.004 \le \delta \kappa_\gamma \le +0.004$, which is moved to about $-0.008 \le \delta \kappa_\gamma \le 0$ for $\lambda_\gamma=0.01$, or moved to $0 \le \delta \kappa_\gamma \le +0.008$ for $\lambda_\gamma=-0.014$. In the case of 1000 GeV center-of-mass energy, the cross-section is not sensitive to the sign of $\lambda_\gamma$, as already discussed. Thus, the deviation of the dependence of the cross-section of $\lambda_\gamma$ from linearity is evidently seen in Fig.~\ref{fig:cs_dkp_lmd} (lower right).  On the other hand, in the case of $\lambda_\gamma$, at 500 GeV with 100 fb$^{-1}$, and keeping $\delta \kappa_\gamma=0$, a limit of $-0.013\le \lambda_\gamma\le +0.01$ could be reached, which is changed to $-0.02\le \lambda_\gamma\le 0.004$ if $\delta \kappa_\gamma=+0.002$, and $-0.005\le \lambda_\gamma\le 0.015$ if $\delta \kappa_\gamma=-0.002$.
 In the case of 1000 GeV center-of-mass energy, the reach on $\lambda_\gamma$ are $-0.013\le \lambda_\gamma\le +0.013$, $-0.008\le \lambda_\gamma\le +0.008$ and $-0.017\le \lambda_\gamma\le +0.017$ corresponding to $\delta \kappa_\gamma=0,~~+0.002$ and $-0.002$, respectively. Thus, quite evidently, the study of the cross-section alone will not provide information regarding the values of or reach of either of the parameters in question.
Possibilities of disentangling these effects will be discussed through a study of different kinematic distributions of the decay products. 

\begin{figure}
\begin{center}
\begin{tabular}{c c}
\hspace{-6mm}
\includegraphics[angle=0,width=88mm]{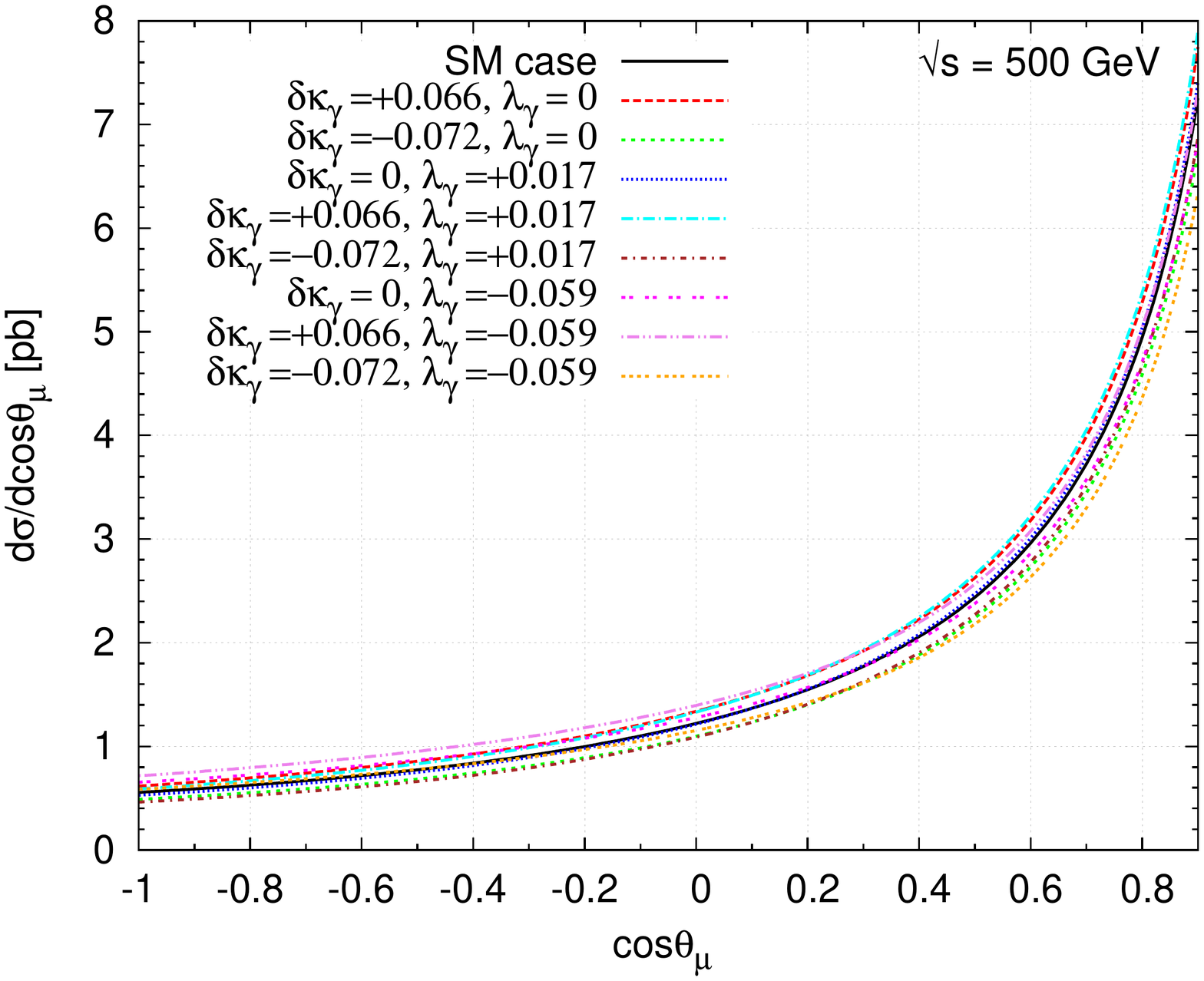}
\hspace{-14mm}
\includegraphics[angle=0,width=88mm]{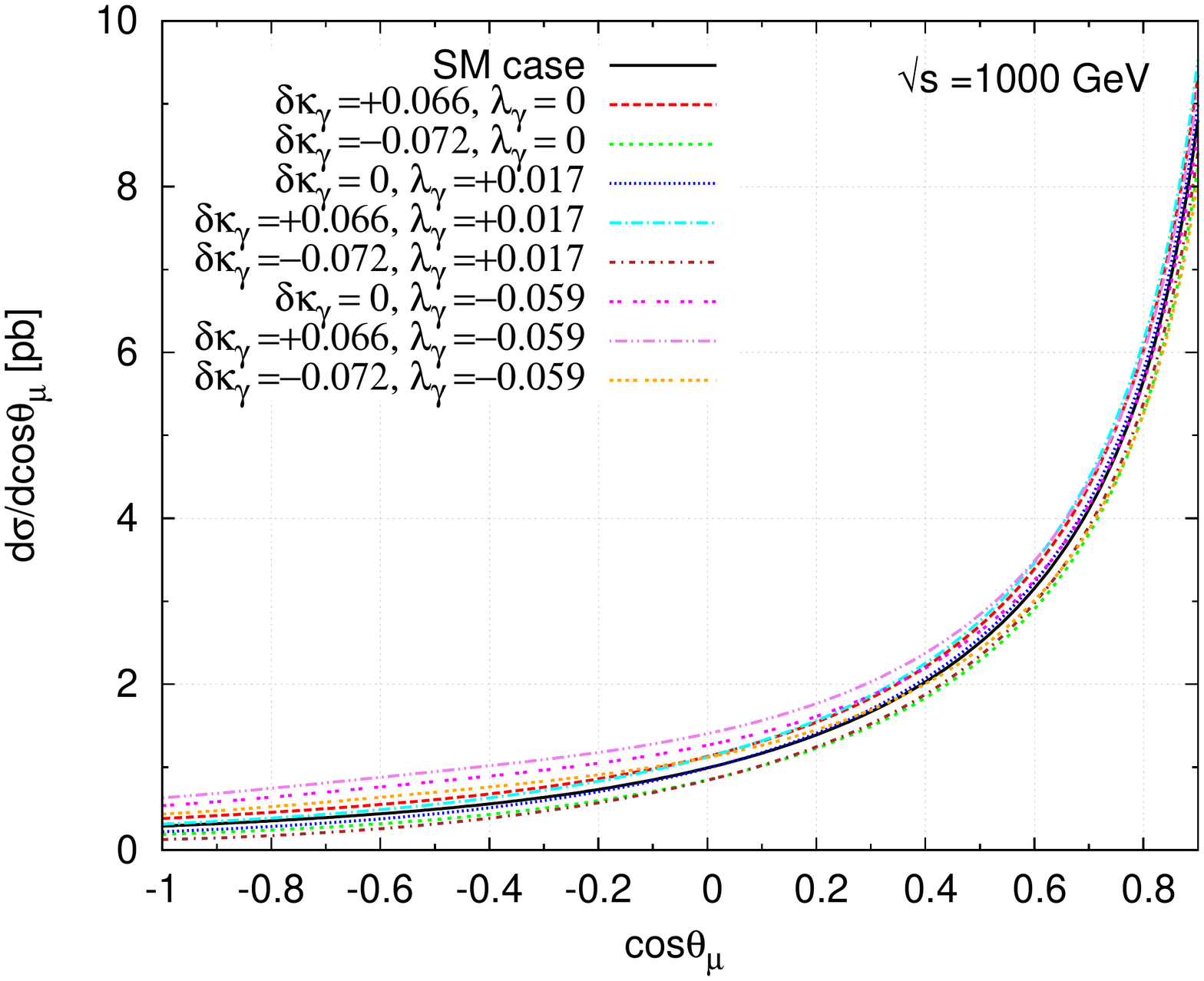}
\end{tabular}
\vspace{-8mm}
\caption{{The angular distribution for different combinations of $\delta\kappa_{\gamma}$ and $\lambda_{\gamma}$ for the two different values of $\sqrt{s}=500$ GeV (left) and $1000$ GeV (right).}}
\label{fig:dsigct1}
\end{center}
\end{figure}

We next consider the angular distribution of the secondary muons plotted in Fig.~\ref{fig:dsigct1} for different combinations of $\delta\kappa_{\gamma}$ and $\lambda_{\gamma}$, at center-of-mass energies of $500$ GeV and $1000$ GeV. As can be seen, the effect is more pronounced at smaller $\cos\theta_\mu$ values, while most of the events are gathered in the forward direction. With high luminosity, one can expect high statistics for single $W$ production at $e\gamma$ collider. This means, that either the entire backward region, or even a smaller region of the $\cos\theta_\mu$ could be probed to see the effect of anomalous TGC.  In Table~\ref{table:AngDist500_1000} the deviation from SM case is illustrated by considering a bin of $-0.90 \le \cos\theta_\mu \le -0.85$ with an integrated luminosity of 100 fb$^{-1}$. The number of events in the backward hemisphere ($N^{\rm back}$), and the deviation from the SM case ($\Delta^{\rm back}=N^{\rm back}-N^{\rm back}_{\rm SM}$) show that deviations as big as 30\% is possible in at $\sqrt{s}=500$ GeV, which is increased to close to 150\% at $\sqrt{s}=1000$ GeV. The deviations are more pronounced in the case of $1000$ GeV, compared to $500$ GeV. The large sensitivity of this observable to the different combinations of the two anomalous couplings gives a way to distinguish different scenarios. A very moderate luminosity of $100$fb$^{-1}$ gives enough statistics to mean that 10\% deviation correspond to 5 - 6 $\sigma$ deviation. Another way to present this effect is through the forward-backward asymmetry, defined as
\begin{eqnarray}
A^{a}_{FB} = \frac{\left[\int_{-1}^{0}\frac{d\sigma}{d\cos\theta_{\mu}}d\cos\theta_{\mu} - \int_{0}^{1}\frac{d\sigma}{d\cos\theta_{\mu}}d\cos\theta_{\mu}\right]}{\left[\int_{-1}^{0}\frac{d\sigma}{d\cos\theta_{\mu}}d\cos\theta_{\mu} + \int_{0}^{1}\frac{d\sigma}{d\cos\theta_{\mu}}d\cos\theta_{\mu}\right]}, \nonumber
\end{eqnarray}

\begin{equation}
\Delta A_{FB}(\%) = \frac{\left|A_{FB}^{Ano.} - A_{FB}^{SM}\right|}{A_{FB}^{SM}}\times100.
\label{eqn:percentageasymmetry}
\end{equation}

\begin{figure}
\begin{center}
\begin{tabular}{c c}
\hspace{-6mm}
\includegraphics[angle=0,width=88mm]{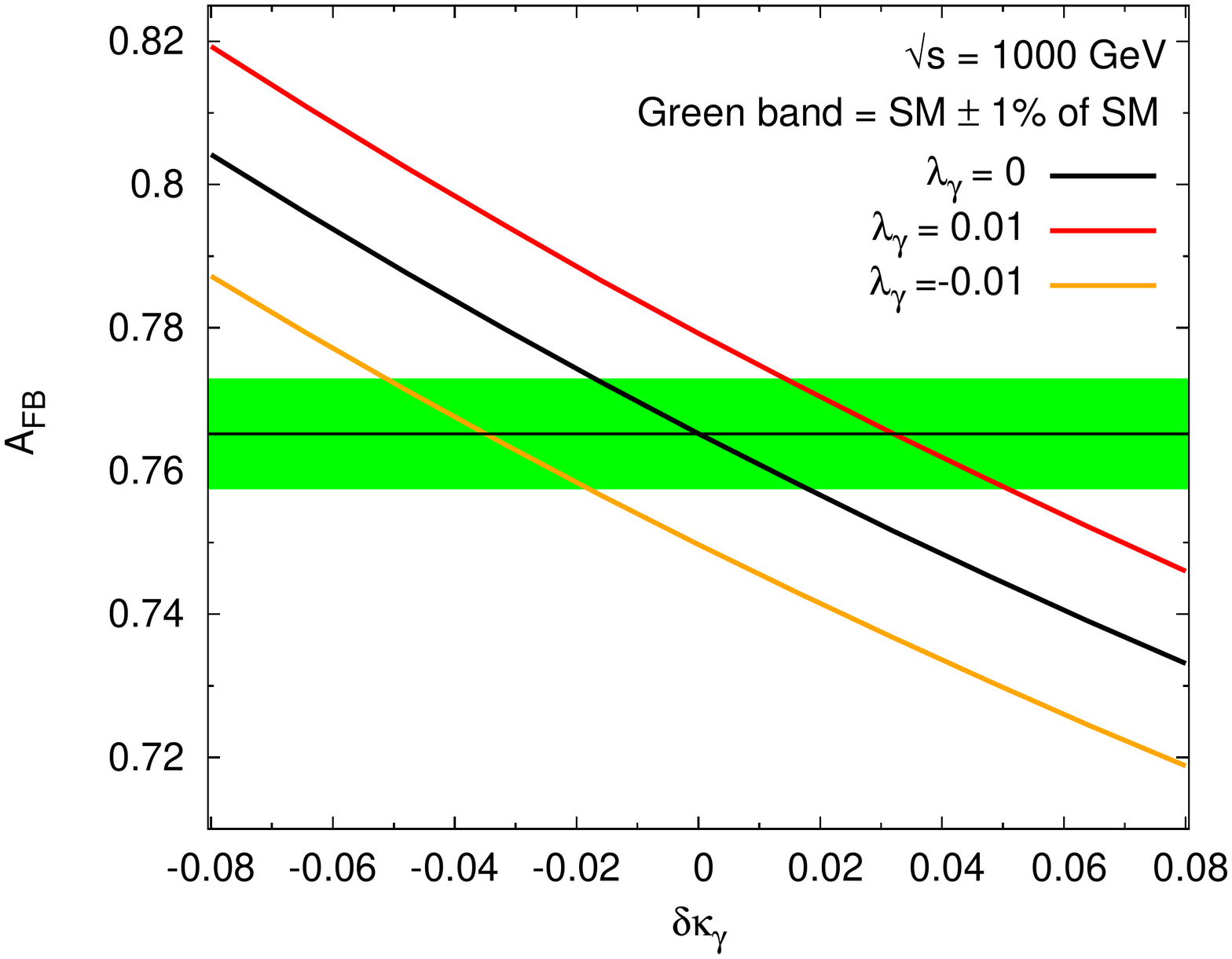}   
\hspace{-14mm}
\includegraphics[angle=0,width=88mm]{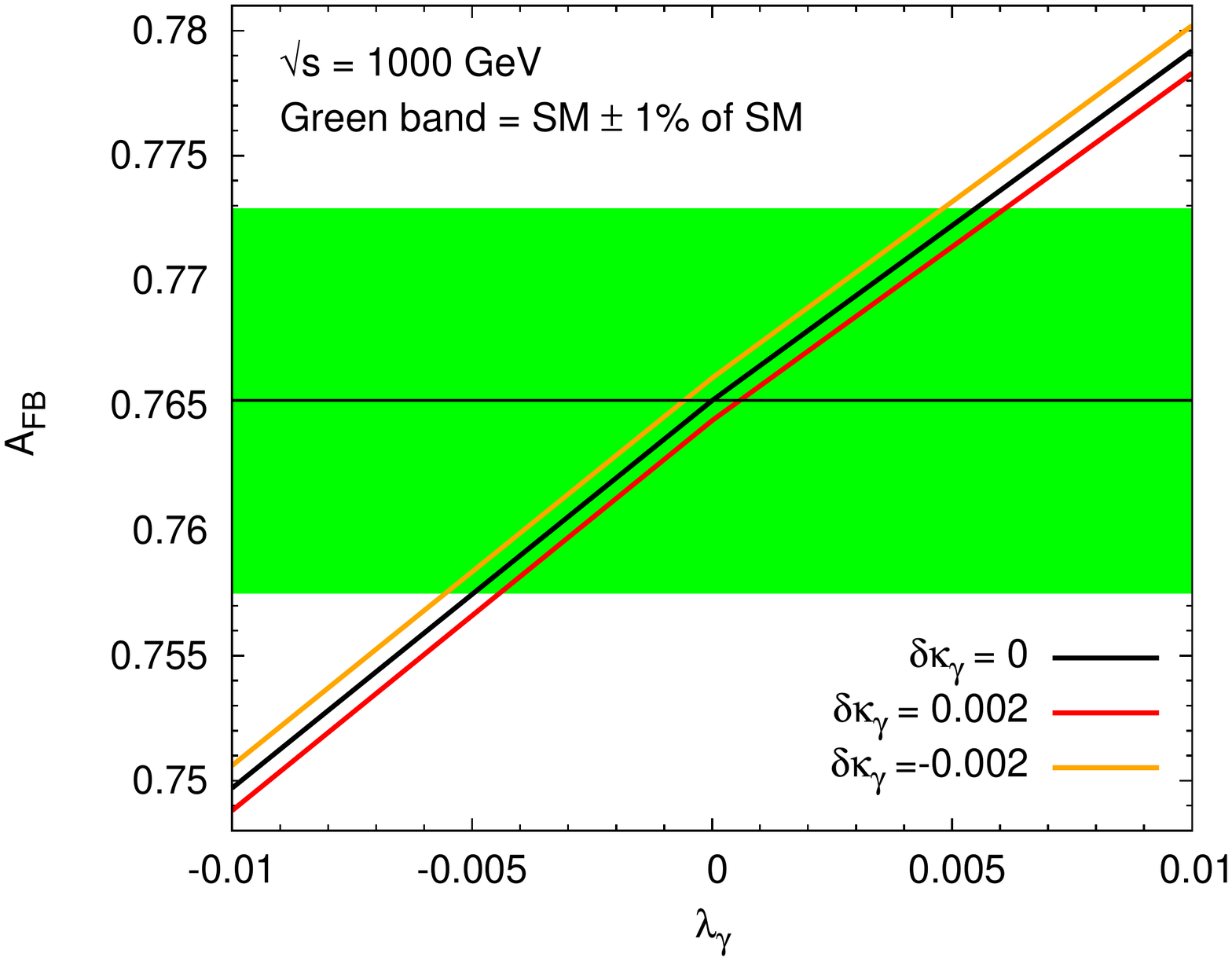} 
\end{tabular}
\vspace{-8mm}
\caption{ {The forward-backward asymmetry against$\delta\kappa_{\gamma}$ (top row) and $\lambda_{\gamma}$ (bottom row) at $\sqrt{s}=500$ GeV (left) and $1000$ GeV (right). The green band correspond to a $\pm 10$\% deviation from the SM case. }}
\label{fig:Asym_FB_dkp_lmd}
\end{center}
\end{figure}

In Fig.~\ref{fig:Asym_FB_dkp_lmd}, this asymmetry is presented for different combinations of the parameter values.  In the SM, at 1000 GeV with 100 fb$^{-1}$ integrated luminosity, we expect about 460000 number of events in total (with 100\% efficiency of detection). Out of this, about 380000 is expected to be in the forward direction giving rise to an asymmetric number of events of about 300000. Considering the statistical uncertainty, even a deviation at a few percent level is perceivable. We have considered a typical 1\% deviation to find the reach of the couplings for this configuration of the collider.
The green band shows the region that falls within 1\% deviation from the SM values. 
The single parameter reach on $\delta \kappa_\gamma$ is better through the consideration of cross-section compared to the forward-backward asymmetry. On the other hand, the case of $\lambda_\gamma$ is quite different. At $\sqrt{s}=1000$ GeV, the reach could be improved by factor of 2 to $-0.005\le \lambda_\gamma \le +0.005$. More importantly, the dependence of the derivable limits, which was about 30\% in the case of the limits drawn from the cross-section, is now reduced to about 10\% at this center-of-mass energy, clearly demonstrating the advantage of the $A_{FB}$ in probing the influence of $\lambda_\gamma$.

\begin{center}
\begin{table}
\begin{tabular}{|cc|c|c|c|c|}
\hline                       
\textbf{$~~\delta\kappa_\gamma~~$} &\textbf{$~~\lambda_\gamma~~$}  &\multicolumn{2}{c|}{$\sqrt{s} = 500$ GeV} &\multicolumn{2}{c|}{$\sqrt{s} = 1000$ GeV} \\\cline{3-6}
             &  &\text{$~~~N^{\rm back}~~~$}  &\text{$~~~~\Delta^{\rm back}~\%~~~~$}         &\text{$~~~N^{\rm back}~~~$}  &\text{$~~~~\Delta^{\rm back}~\%~~~~$}    \\\cline{1-6} 
\hline\hline
\multicolumn{2}{|c|}{\rm SM case}&2900  & 0   & 1600   & 0   \\\cline{1-6}
\hline\hline
 0.066 & 0       &3250  & 12  & 2100   & 31  \\\cline{1-6}
-0.072 & 0       &2600  & $-10$  & 1100   & $-31$  \\\cline{1-6}
 0     & 0.017   &2800  & $-3$   & 1250   & $-22$  \\\cline{1-6}
 0     &-0.059   &3400  & 17  & 2850   & 78  \\\cline{1-6}
 0.066 & 0.017   &3100  & 7   & 1700   & 6   \\\cline{1-6}
 0.066 &-0.059   &3750  & 30  & 3900   & 144 \\\cline{1-6}
-0.072 & 0.017   &2400  & $-17$  & 750    &$-53$  \\\cline{1-6}
-0.072 &-0.059   &3050  & 5   & 2350   & 43  \\\cline{1-6}
\hline
\end{tabular}
\caption{{The number of events within $-0.90\le \cos\theta_{\mu} \le-0.85$ for different combinations of $\delta\kappa_{\gamma}$ and $\lambda_{\gamma}$ at $\sqrt{s}=500$ GeV and $1000$ GeV, along with the corresponding deviation from the SM case. An integrated luminosity of $100~fb^{-1}$ is considered.}}
\label{table:AngDist500_1000}
\end{table}
\end{center}

\begin{figure}
\begin{center}
\begin{tabular}{c c}
\hspace{-6mm}
\includegraphics[angle=0,width=88mm]{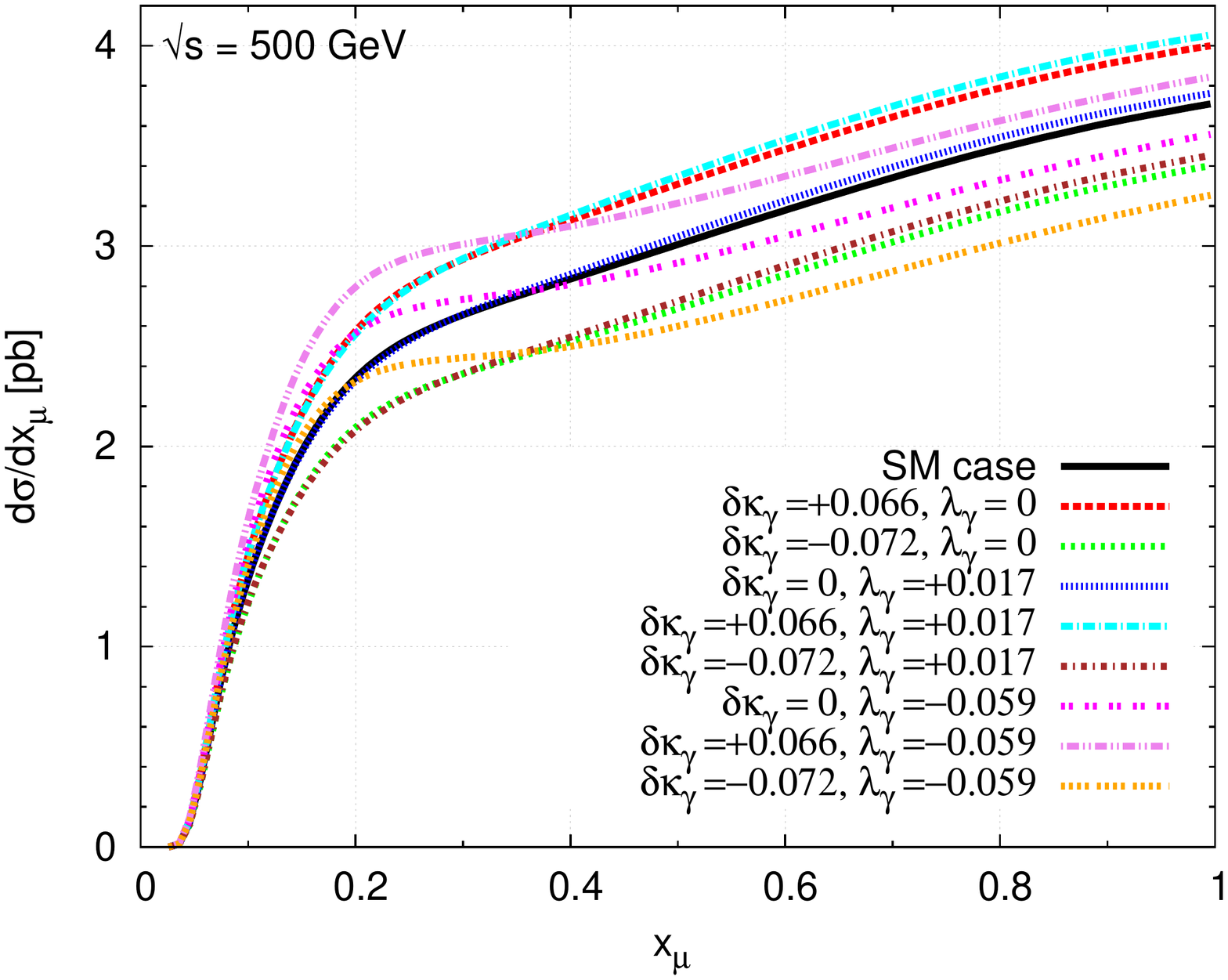}
\hspace{-14mm} 
\includegraphics[angle=0,width=88mm]{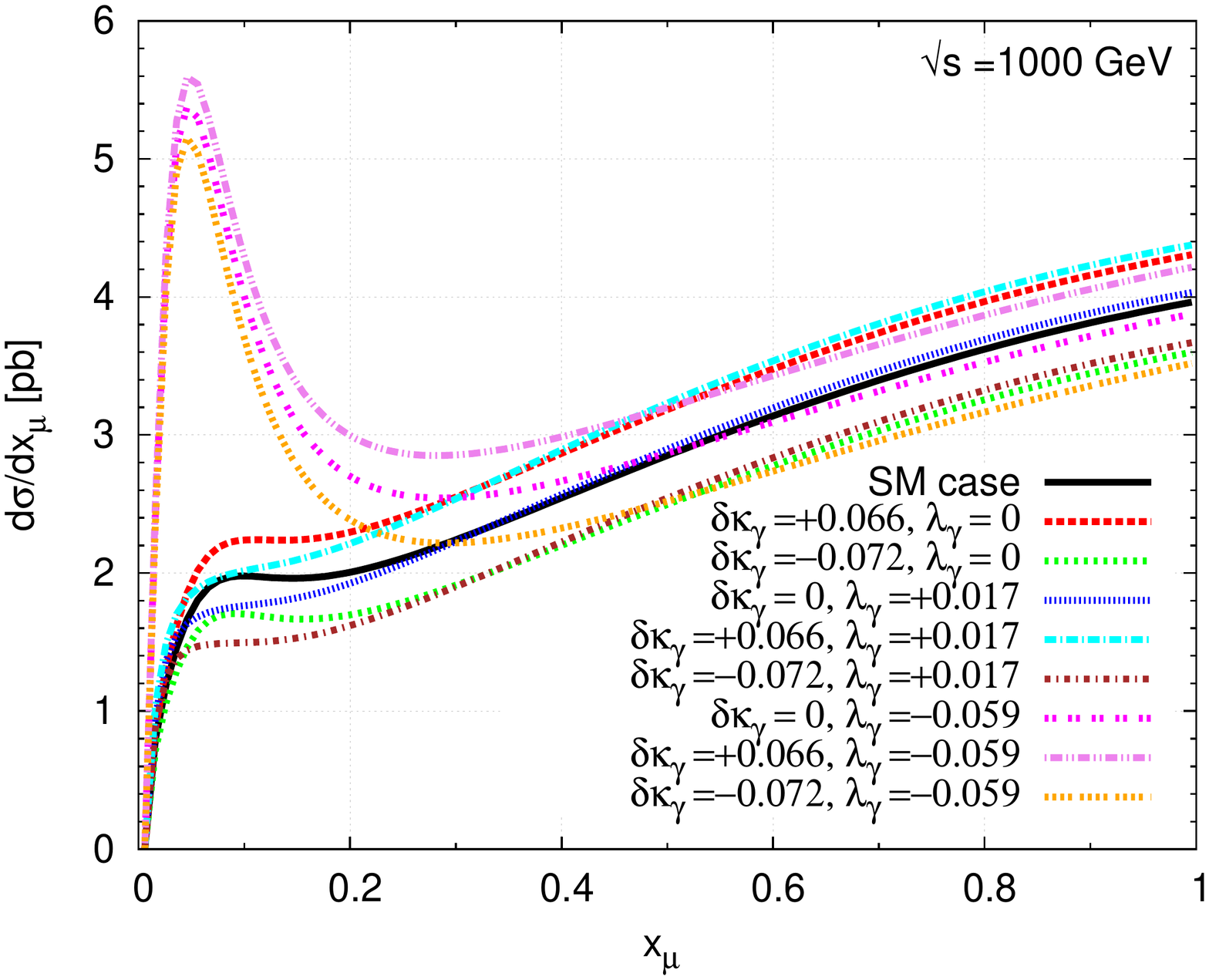} 
\end{tabular}  
\vspace{-8mm}
\caption{ { The energy distribution of the secondary muon for different combinations of $\delta\kappa_{\gamma}$ and $\lambda_{\gamma}$, compared with the SM case. Center-of-mass energies considered are  $\sqrt{s}=500$ GeV (left) and $1000$ GeV (right).}}
\label{fig:energy_dist90}
\end{center}
\end{figure}

So far, we have noted the influence of one of the couplings in deriving the bound on the other. In order to see the possibility of disentangling the effects of these couplings, we shall turn to the kinematic distributions like the energy distribution, alone and in combination with the angular distribution. In Fig.\ref{fig:energy_dist90}, the energy distribution against $x_\mu=\frac{E_\mu}{E_e}$, where $E_e$ is the electron beam energy is presented, showing the effect of anomalous couplings for different combination of TGC parameters at $\sqrt{s}$ of $500$ GeV and $1000$ GeV.  Once again, we emphasis that it is only through a combination of different observables that one may be able to disentangle information regarding the $\delta \kappa_\gamma$ and $\lambda_\gamma$ couplings. The influence of $\delta \kappa_\gamma$ remains more or less the same throughout the range of $x_\mu$. At the same time, $\lambda_\gamma > 0$ has a slight diminishing effect in the low energy region, and a slight enlarging effect at the high energy region. On the other hand, $\lambda_\gamma < 0$ has a large enlarging effect at low energies , which turns the other way around, but with a much smaller magnitude at higher energies. Thus, the energy distribution of the secondary muons become a clear discriminator of these qualitatively different scenarios, as summarized in Table~\ref{table:energy}.

\begin{center}
\begin{table}
\begin{tabular}{| l | l |}
\hline                       
 Scenario&Effect on $\mu$ energy \\\cline{1-2} 
\hline\hline
&\\[-3mm]
 $\delta \kappa_\gamma > 0$& enhancing effect \\
$\lambda_\gamma \ge 0$& both low and high energies\\ [1mm]\cline{1-2}
&\\ [-3mm]
 $\delta \kappa_\gamma < 0$& diminishing effect \\
$\lambda_\gamma \le 0$& both low and high energies \\ [2mm] \hline
&\\[-3mm]
 $\delta \kappa_\gamma  \ge  0 $& large enhancing effect at low energies \\
$\lambda_\gamma<0$&diminishing effect at high energies, depending on $\delta \kappa_\gamma$\\ [2mm] \hline
&\\[-3mm]
 $\delta \kappa_\gamma < 0$& large enhancing effect at low energies \\
$\lambda_\gamma<0$& small enhancing effect at high energies, depending on $\delta \kappa_\gamma$\\ [2mm] \hline
\end{tabular}
\caption{Disentangling different scenarios based on the energy distribution of the secondary $\mu$.}
\label{table:energy}
\end{table}
\end{center}

\begin{figure}
\begin{center}
\begin{tabular}{c c}
\includegraphics[angle=0,width=60mm]{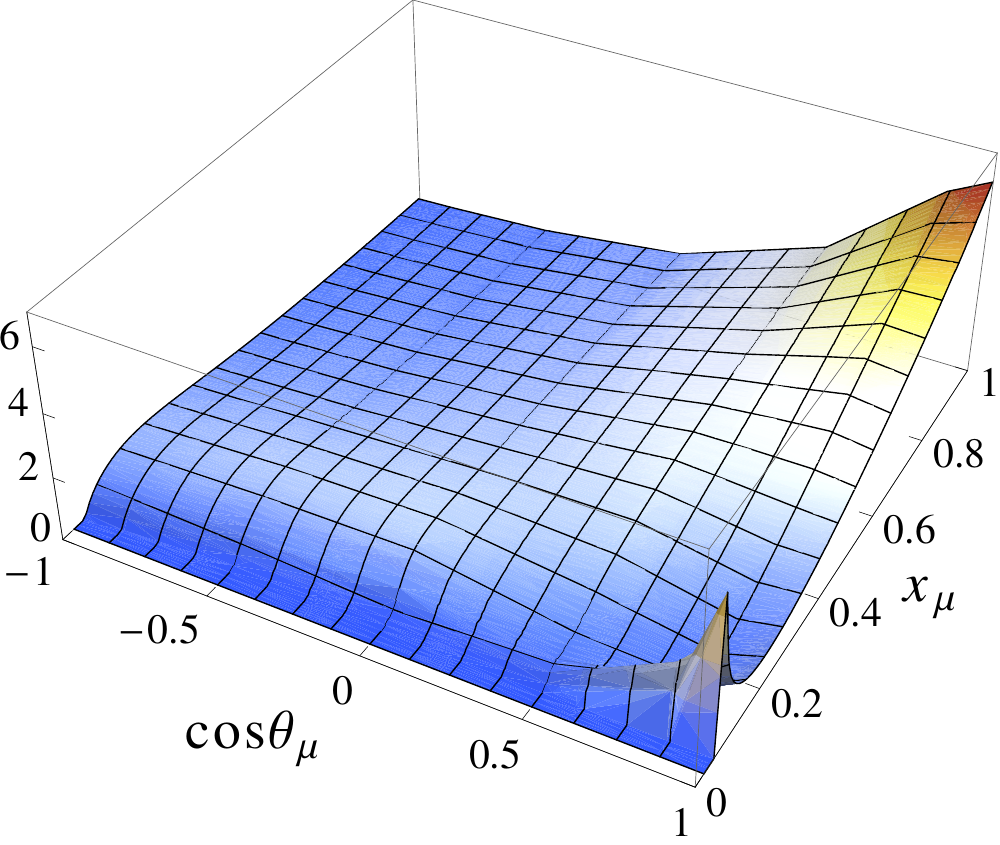}&
\hspace{14mm} 
\includegraphics[angle=0,width=60mm]{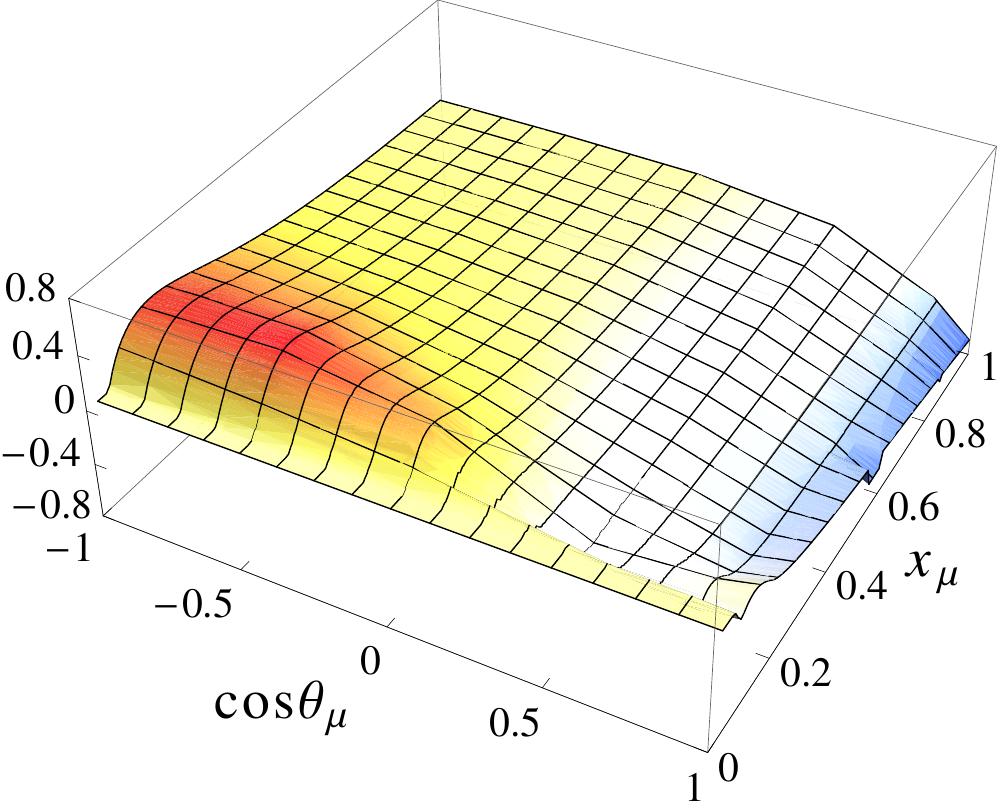} \\
\includegraphics[angle=0,width=60mm]{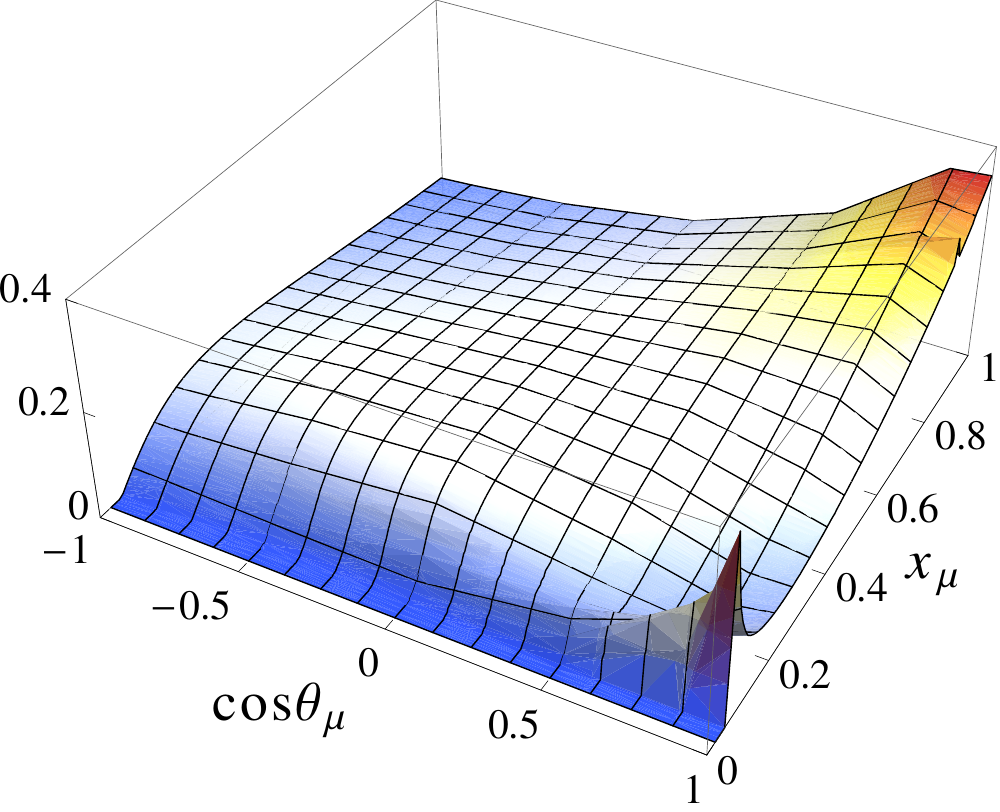} &
\hspace{14mm} 
\includegraphics[angle=0,width=60mm]{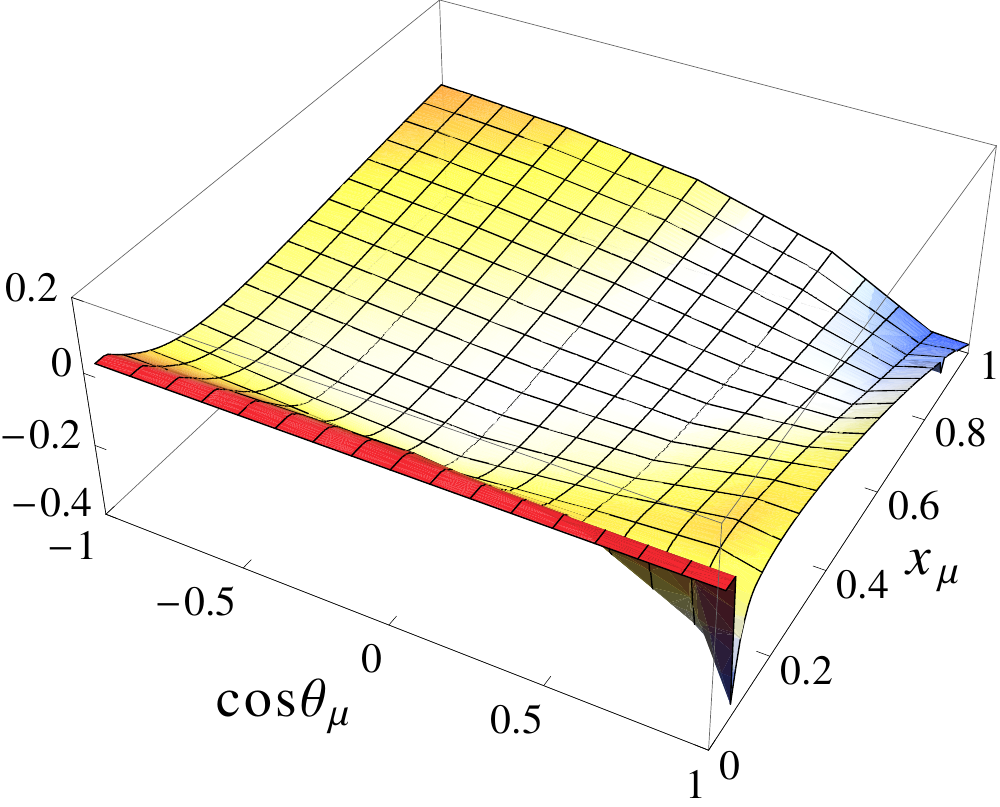} 
\end{tabular}  
\vspace{0mm}
\caption{ { The energy-angle double distribution of the secondary muon (along the vertical axis: $\frac{d\sigma}{dx_\mu ~d\cos\theta_\mu}$ (pb)) at $\sqrt{s}=500$ GeV. Top left is the SM case. Other cases are deviations from the SM; $\delta \kappa_\gamma=0,~~\lambda_\gamma=-0.059$ (top right);   $\delta \kappa_\gamma=+0.069,~~~\lambda_\gamma=0$ (bottom left);  $\delta \kappa_\gamma=-0.072,~~~\lambda_\gamma=0$ (bottom right).}}
\label{fig:energy-angle}
\end{center}
\end{figure}

A more detailed picture could be made available by exploiting the energy-angle double distribution, as presented in Fig.~\ref{fig:energy-angle}, where the SM distribution (top left) and the deviations from the SM in the presence of $\delta \kappa_\gamma$ and $\lambda_\gamma$, individually,  are shown. Deviation in the case of $\lambda_\gamma>0$ is negligible, and therefore not presented.  Evidently, in order to probe $\lambda_\gamma$, one need to focus on the low energy muon emerging in the backward region, with with some trace in the high energy - backward region, and very low energy forward region. On the other hand, the signature of $\delta \kappa_\gamma$ will be seen more visibly in the forward direction for high energy $\mu$. In the case of low energy $\mu$, the effect of $\delta \kappa_\gamma$ is somewhat independent of the angle, except for the extreme forward region. The reason for these complex dependence is a result of the complex interdependence of the kinematic variables and the parameters, the origin of which are difficult to decipher,.

\begin{figure}
\begin{center}
\begin{tabular}{c c}
\hspace{-6mm}
\includegraphics[angle=0,width=88mm]{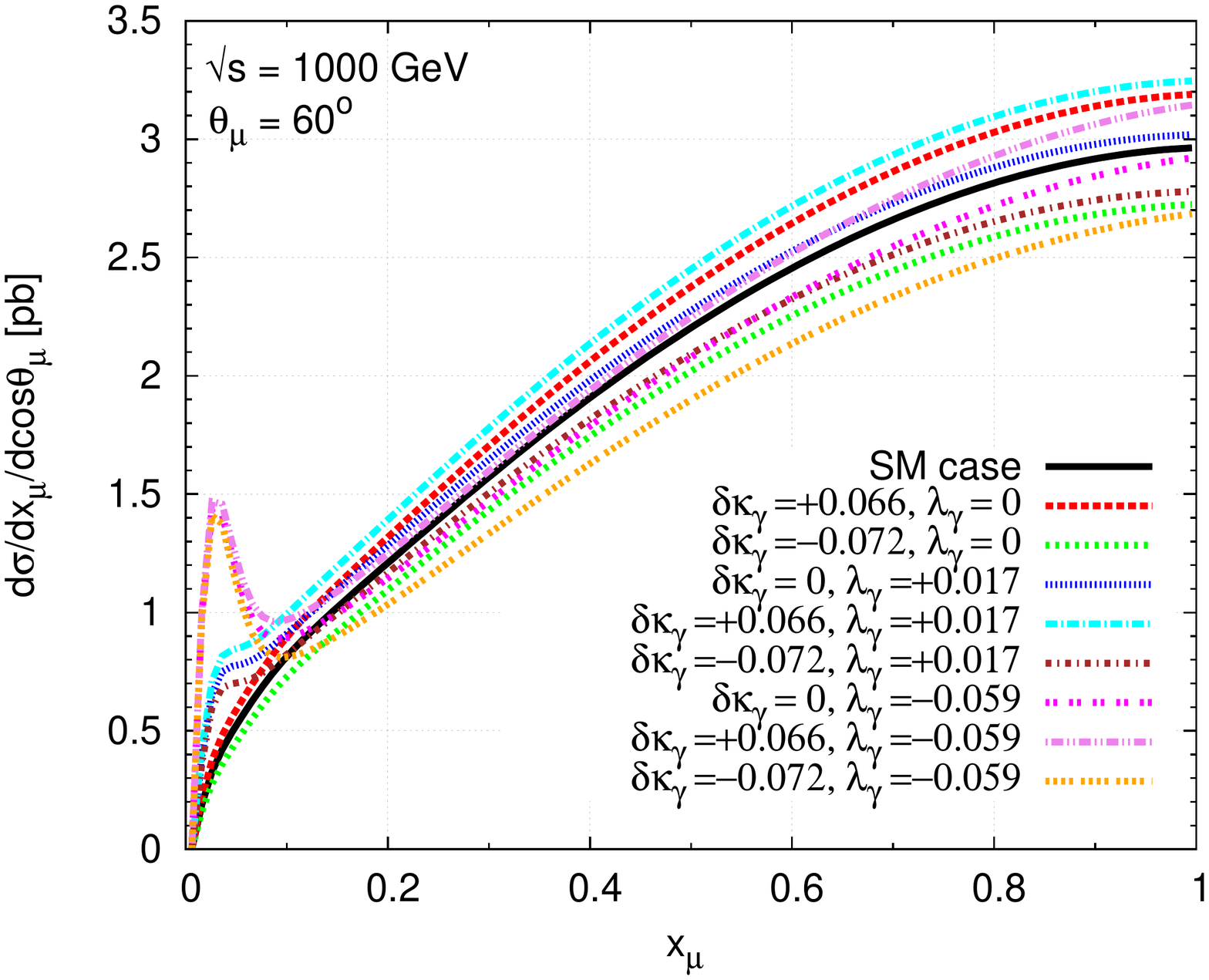} 
\hspace{-14mm}
\includegraphics[angle=0,width=88mm]{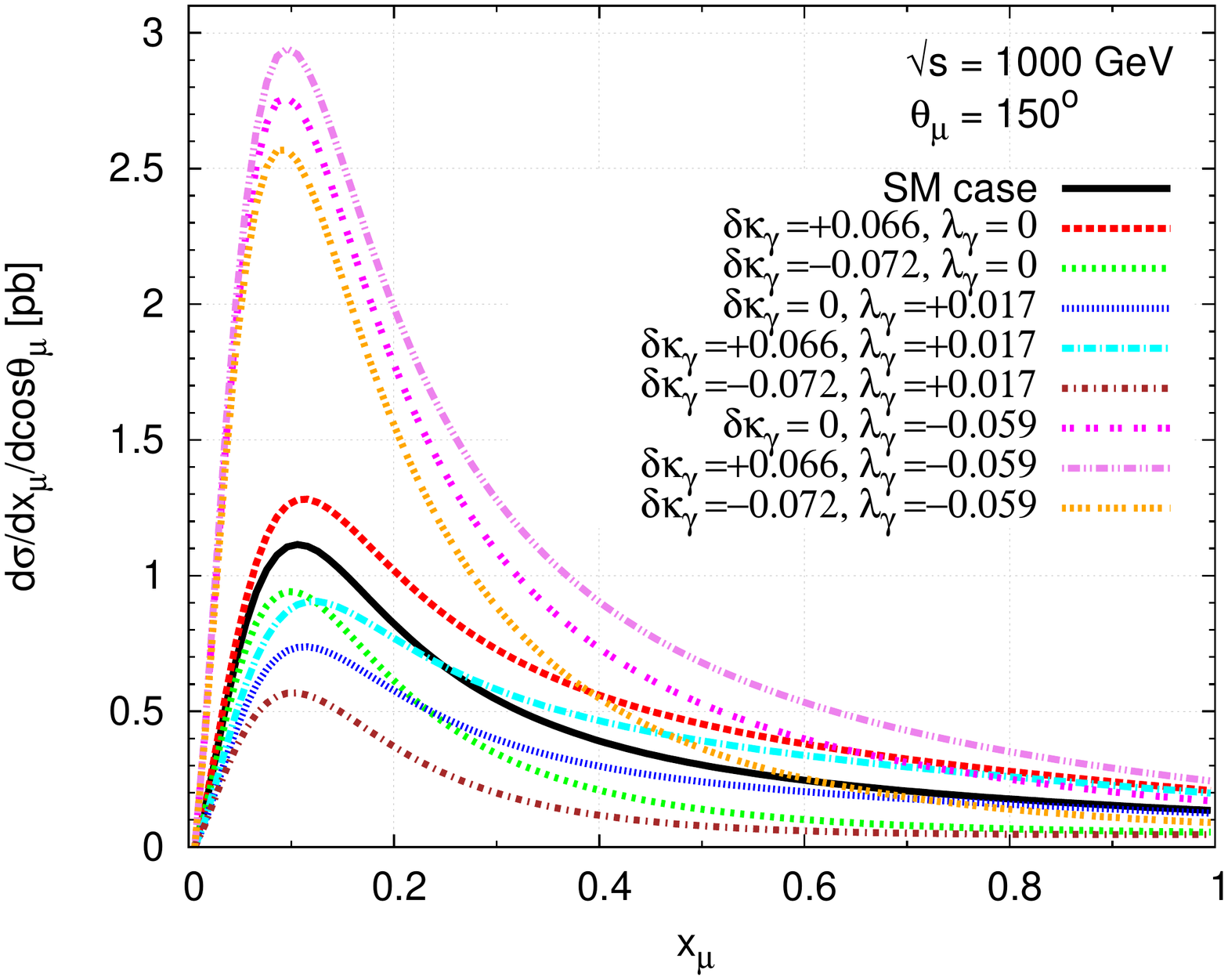}  
\end{tabular}
\vspace{-8mm}
\caption{{The energy distribution of the secondary muon for different combinations of $\delta\kappa_{\gamma}$ and $\lambda_{\gamma}$, considered at muon angles of $\theta_{\mu} = 60^o$ and $150^o$, at  $\sqrt{s}=1000$ GeV.}}
\label{fig:energy_ang_1000}
\end{center}
\end{figure}

Taking slices from the above surface plots, the  energy distributions at fixed angles of $60^o$ and $150^o$ are presented
in Fig.~\ref{fig:energy_ang_1000} corresponding to center-of-mass energy of $1000$ GeV. These angles represent the typical behaviour in the low and high angle regions, respectively. Quantifying these effects, the number of events in the chosen energy bins for the above angles, corresponding to the SM case as well as different combinations of the anomalous couplings, along with the percentage deviations is given in Table~\ref{table:EngAngDist}.  The energy bins are chosen to maximize the effect, in each case. In the forward region with high energy muon, where the effect of $\lambda_\gamma$ is small, whereas the $\delta \kappa_\gamma$ has large effects has a maximum of 12\% in the case of $\sqrt{s}=500$ GeV, which remains more or less at the same level (10\%) in the case of $\sqrt{s}=1000$ GeV. On the other hand, the presence of $\lambda_\gamma$ has much larger impact, with 33\% when $\delta \kappa_\gamma$ is zero at $\sqrt{s}=500$ GeV, which is increased to 41\% along with positive values of $\delta \kappa_\gamma$ and decreased to 21\% for its negative values. These effects can be increased to more than 4 times for $\sqrt{s}=1000$ GeV. 

\begin{center}
\begin{table}[tp!]
\resizebox{\textwidth}{3.0cm}{ 
\begin{tabular}{|cc|c|c|c|c|c|c|c|c|}
\hline
&&\multicolumn{4}{c|}{$\sqrt{s} = 500$ GeV}&\multicolumn{4}{c|}{$\sqrt{s} = 1000$ GeV}\\\cline{3-10}
& &\multicolumn{2}{c|}{$\theta_\mu = 60^o$} &\multicolumn{2}{c|}{$\theta_\mu = 150^o$} &\multicolumn{2}{c|}{$\theta_\mu = 60^o$} &\multicolumn{2}{c|}{$\theta_\mu = 150^o$}    \\
   &   &\multicolumn{2}{c|}{$x_\mu=0.95-1.0$} &\multicolumn{2}{c|}{$x_\mu=0.056-0.106$} &\multicolumn{2}{c|}{$x_\mu=0.95-1.0$} &\multicolumn{2}{c|}{$x_\mu=0.056-0.106$}  \\\cline{3-10}
   \textbf{$\delta\kappa_\gamma$} &\textbf{$\lambda_\gamma$}   &\text{$~N_{events}~$}  &\text{$~~\Delta~[\%]~~$}  &\text{$~N_{events}~$}  &\text{$~~\Delta~[\%]~~$} &\text{$~N_{events}~$}  &\text{$~~\Delta~[\%]~~$}  &\text{$~N_{events}~$}  &\text{$~~\Delta~[\%]~~$}    \\\cline{1-10}
\hline\hline
\multicolumn{2}{|c|}{SM Case}&13500 & --    &6200  & -- &14750 &0        &6000   &0   \\\cline{1-10}
\hline\hline
 0.066 & 0          &14550 &8       &6750  &$9$ &15900 &8        &7000    &17 \\\cline{1-10}
-0.072 & 0          &12450 &$-8$  &5650  &$-9$ &13600 &$-8$  &5000 &$-17$ \\\cline{1-10}
 0     & 0.017      &13750 &2        &5750  &$-7$ &15000 &2        &4000    &$-33$ \\\cline{1-10}  
 0     &-0.059      &13000 &$-4$   &8250  &33 &14500 &$-2$  &14500  &141   \\\cline{1-10}
 0.066 & 0.017   &14750 &9        &6250  &1 &16225 &10      &4750    &$-21$   \\\cline{1-10}
 0.066 &-0.059   &14000 &4        &8750  &41&15600 &6        &15500  &158  \\\cline{1-10}
-0.072 & 0.017  &12650 &$-6$   &5250  &$-15$ &13850 &$-6$   &3000    &$-50$  \\\cline{1-10}
-0.072 &-0.059   &11900&$-12$ &7500  &21  &13300 &$-10$ &13400  &123 \\\cline{1-10}
\hline
\end{tabular}
}
\caption{ {The number of events within specified $x_\mu$ bins, and for fixed $\theta_\mu$, for different combinations of $\delta\kappa_{\gamma}$ and $\lambda_{\gamma}$ at $\sqrt{s}=500$ GeV and $1000$ GeV, along with the corresponding percentage deviation ($\Delta$) from the SM case.  An integrated luminosity of $100$fb$^{-1}$ is considered.}}
\label{table:EngAngDist}
\end{table}
\end{center}

\begin{figure}
\begin{center}
\begin{tabular}{c c}  
\hspace{-6mm}
\includegraphics[angle=0,width=88mm]{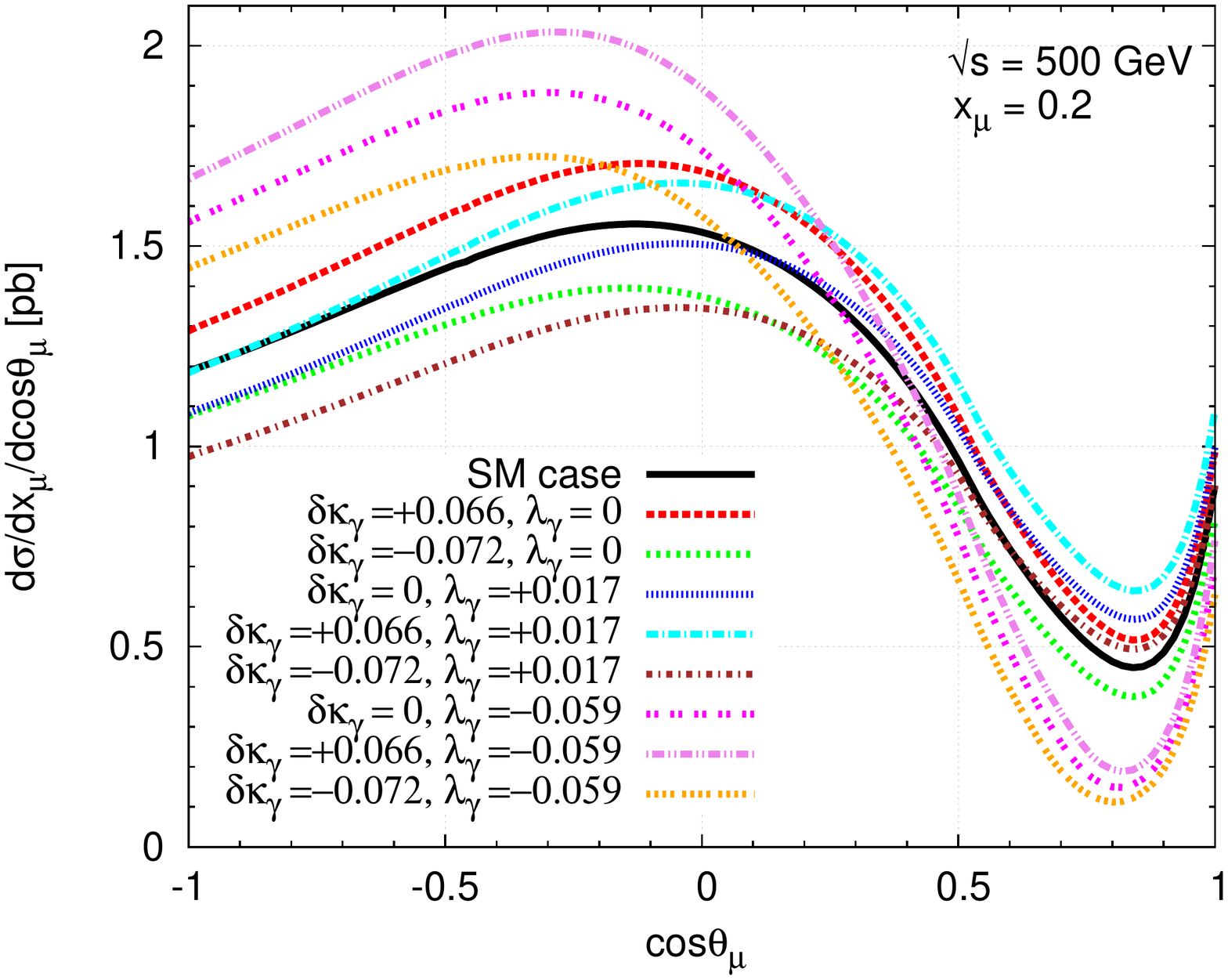} 
\hspace{-14mm}
\includegraphics[angle=0,width=88mm]{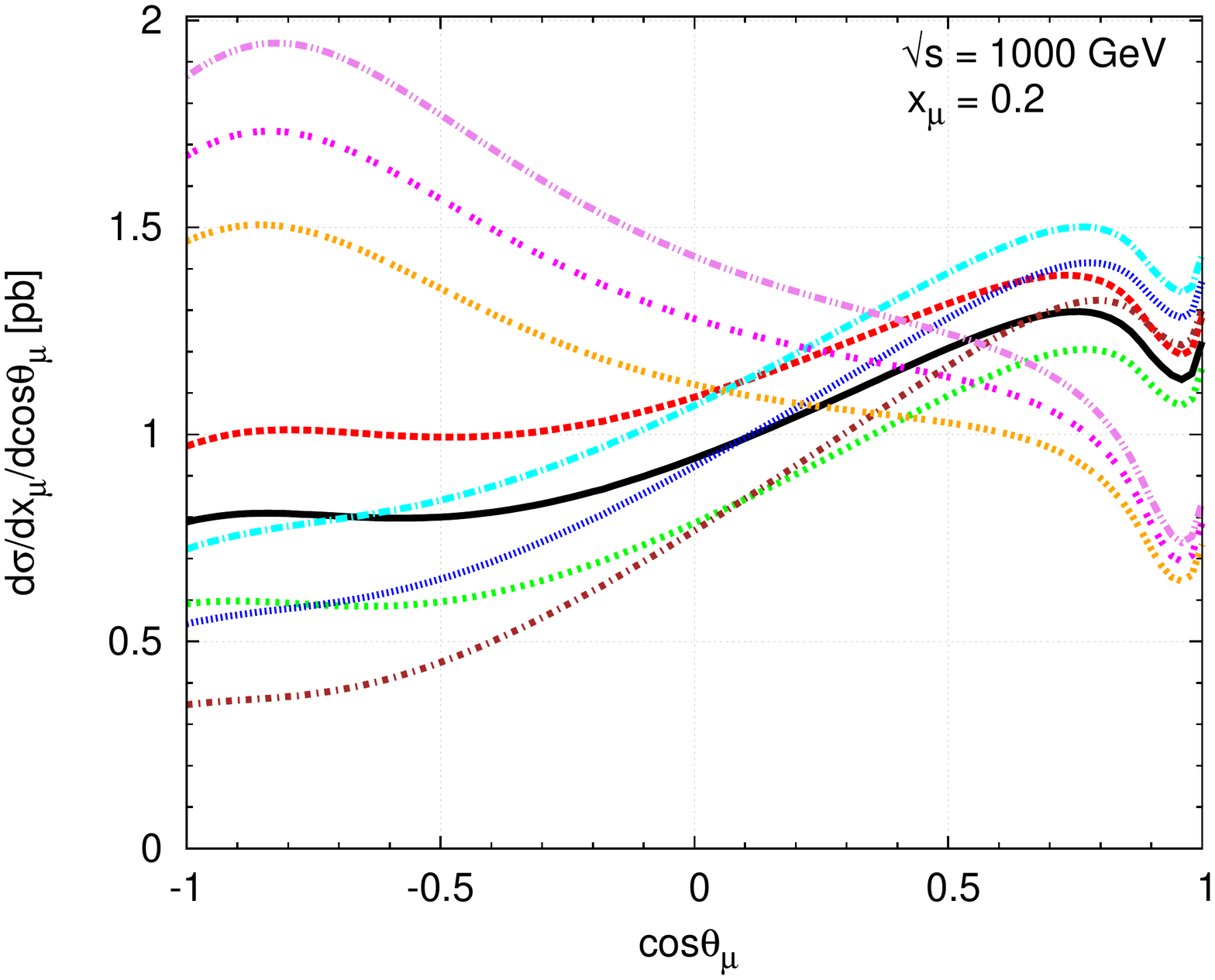}\\
\end{tabular}
\vspace{-8mm}
\caption{{The energy-angle distribution of the secondary muon for different combinations of $\delta\kappa_{\gamma}$ and $\lambda_{\gamma}$, considered at fixed muon energy parameter ($x_{\mu}$) at $\sqrt{s}$ of $500$ GeV and $1000$ GeV. Colour coding is the same in all plots.}}
\label{fig:energyangle_dist_xp}
\end{center}
\end{figure}

Focusing on the low energy region, we consider variation with $\theta_\mu$ for fixed $x_\mu=0.2$ (as a typical value), and present difference scenarios in Fig.~\ref{fig:energyangle_dist_xp}. While the absence of $\lambda_\gamma$ and its positive values for different $\delta \kappa_\gamma$ values have moderate impact, especially in the backward region, the negative values of $\lambda_\gamma$ could produce significantly different distributions. In order to quantify the effects, we have computed the forward-backward asymmetry in each of these cases, and tabulated in Table~\ref{table:asymmetryZA1000}. The deviations are evidently more pronounced in all cases, and  much larger in the $\lambda_\gamma \le 0$ case. Note that owing to large cross-section, even limiting to a small energy bin of $\Delta x_\mu=0.2$ around the $\mu$ energy considered will lead to about 40000 events, when integrated over the angle.  Thus, a change in the forward-backward asymmetry to the tune of 200\% is certainly detectable. Coming to the reach on the value of the couplings achieved through these observations, we note that the asymmetry in the SM case is about $-19\%$, and $+40\%$ at $\sqrt{s}=500$ GeV and 1000 GeV, respectively. We summarize the extend of asymmetry measurement  in Table~\ref{table:reach}, considering purely statistical uncertainty. This suggests that deviations from the asymmetry of about 5 \% could be expected at $3\sigma$ level for . Reading it along side Table~\ref{table:asymmetryZA1000} indicates a reach of about an order of magnitude better than that the value considered there, which means about $\lambda_\gamma \sim - 0.003$, assuming a linear dependence.

\begin{table}
\begin{center}
\begin{tabular}{|rr|c|c|}
\hline
\multicolumn{2}{|c|}{${\delta \kappa_\gamma~~~~~~~~~\lambda_\gamma}$} &\multicolumn{2}{c|}{\textbf{$\Delta A_{FB}(\%)$}}  \\\cline{3-4}
\multicolumn{2}{|c|}{ } &$\sqrt{s} = 500$ GeV & $\sqrt{s} = 1000$ GeV \\
\hline\hline
$0.066$& $ 0$   &$-4$ &$-33$       \\\cline{1-3}
\hline
$-0.072$&$0$    &5 &$ 48$       \\\cline{1-3}
\hline
$0$&$0.017$   &$-34$ &$ 72$       \\\cline{1-3}
\hline
$0$&$-0.059$   &$92$ &$-205$       \\\cline{1-3}
\hline
$0.066$&$ 0.017$   &$-36$ &$29$       \\\cline{1-3}
\hline
$0.066$&$-0.059$   &$83$ &$-213$       \\\cline{1-3}
\hline
$-0.072$&$0.017$   &-32&$135$       \\\cline{1-3}
\hline
$-0.072$&$-0.059$   &102 &$-194$       \\\cline{1-3}
\hline
\hline
\multicolumn{2}{|c|}{SM~case;  $A_{FB}=$}    &$-0.185$&$0.398$    \\\cline{1-3}
\hline
\end{tabular}
\caption{ {Observed forward-backward asymmetry and its deviation from the SM with a fixed muon energy corresponding to $x_\mu=0.2$, at  $\sqrt{s}=500$ and $1000$ GeV.}}
\label{table:asymmetryZA1000}
\end{center}
\end{table}

\begin{table}
\begin{center}
\begin{tabular}{|c|c|c|c|c|c|}
\hline
&&&&&\\
&$N^F$&$N^B$&$A^{FB}$ \%&$\Delta A^{FB \%}$ &reach on $\lambda_\gamma$\\
&&&&&at $(3\times \Delta A^{FB})$ level\\ \cline{1-6}
SM&18000&26000&18.5&&\\
$N^i\pm \sqrt{N^i}$&[17866,18134]&[25839,26161]&[17.5,18.8]&[-5.0,+1.6]&-0.003\\
\hline
\end{tabular}
\caption{ {Number of events in the forward ($N^F$) and backward ($N^B$) hemispheres, and the corresponding forward-backward asymmetry ($A^{FB}$) in the case of the SM with a fixed muon energy of 50 GeV (corresponding to $x_\mu=0.2$) considering a bin of $\Delta x_\mu=0.2$  ($\Delta E_\mu=50$ GeV), at  $\sqrt{s}=500$ GeV and luminosity of 100 fb$^{-1}$.}}
\label{table:reach}
\end{center}
\end{table}
\section{Summary and Conclusions}

With the $W$ decaying into muons, tthe single $W$ production in $e\gamma$ collisions has a spectacular final state of one single muon, and missing energy. The process is well suited to study the $WW\gamma$ coupling. It may be noted that the process has the advantage that, unlike other relevant process like the $W$ pair production at $e^+e^-$ collisions, it is devoid of effects from other anomalous couplings like $WWZ$. Considering this process, angular distribution of secondary muons in the lab frame is easily constructed, and is readily available in the literature. At the same time, such is not the case of its energy distribution. Here we have presented a semi-analytical way to explore the secondary lepton energy-angle distributions in $e\gamma\rightarrow \nu W$ with $W\rightarrow l\bar \nu$. The advantage of such an observable in analyzing the SM case and probing possible new physics effects is demonstrated.  Variables being defined in the lab frame, are directly used to probe different kinematic regions, so as to explore the sensitivity of the anomalous couplings.
 
 We have derived possible limits on the anomalous couplings $\delta \kappa_\gamma$ and $\lambda_\gamma$, which may be obtained from the cross-section measurements. It is shown that, assuming the other couplings is absent, $3\sigma$ level of cross-section can probe $\delta \kappa_\gamma$ to the level of $\pm 0.004$, and constrain $\lambda_\gamma$ to $-0.013 \le \lambda_\gamma\le+0.01$, for a moderate luminosity of 100 fb$^{-1}$ at a center-of-mass energy of $500$ GeV. The large influence of each of these couplings on the sensitivity of the other is quite visible in the derivable limits. Scanning the two parameter plane the $3\sigma$ bound is presented, showing that $\delta \kappa_\gamma$ could be bounded to small regions for particular values of $\lambda_\gamma$.  The angular distributions along with forward-backward asymmetry improve the situation with better sensitivity. With the large cross-section, the statistics is very good even for 100 fb$^{-1}$ integrated luminosity, which could measure the forward-backward asymmetry to a few percent level. It is found that this could be utilized to improve the limits on $\lambda_\gamma$, whereas not so in the case of $\delta \kappa_\gamma$.  
Exploring different kinematic regions, and studying suitably constructed observables would be able to distinguish different scenarios involving the couplings. We studied the energy and angle distributions of the $\mu$ separately and together to understand how these could be fruitfully employed for this purpose. The energy-angle double distribution enables one to identify the kinematic regions sensitive to different couplings, and their signs. We have found that focusing on the low energy $\mu$ in the backward direction are more sensitive to the case of $\lambda_\gamma <0$, whereas the $\delta \kappa_\gamma$ does not show distinguishable preference to the kinematic regions.
Our analysis shows that the case of (i) $\lambda=0$, or non-zero and positive, (ii) non-zero and negative can be distinguished this way, as explained in Table~\ref{table:energy}. On the other hand, the sign of $\delta \kappa_\gamma$ can be distinguished in the absence of $\lambda_\gamma$, whereas it is hard to do this if $\lambda_\gamma$ is non-zero. At $\sqrt{s}=500$ GeV, with 100 fb$^{-1}$ luminosity, the bound on the $\lambda_\gamma$, when it is negative could be improved by about a factor of 2 by considering the forward-backward asymmetry at a fixed energy ($E_\mu$), compared to the case when the energy is integrated out.

We are aware that the present considerations of the ILC do not favour a photon collider option as priority. At the same time, we hope that this study will help the case for this option at ILC, and expect it to be available eventually. Meanwhile, although more accurate estimates incorporating collider and detector effects, and beam polarization effects are needed to make more quantitative conclusions, we hope our analyses has shown that the leptonic decay of the $W$ produced in $e\gamma$ collision is one of the neat and simple process to probe the $WW\gamma$ coupling without the complications of other couplings. Further, the process has the potential to probe the couplings to a few per-mil level with suitably constructed observables, which could also be used to distinguish different scenarios involving the two parameters, $\delta \kappa_\gamma$ and $\lambda_\gamma$.

\vskip 5mm
\noindent
{\Large \bf Acknowledgment}: \\[2mm]
 PP's work is partly supported by a BRNS, DAE, Government of India (Project No.: 2010/37P/49/BRNS/1446).

\vskip 5mm
\noindent
 {\Large \bf Appendix: Photon luminosity distribution}\\[2mm]
The colliding photons in a realistic electron-photon collider does not have a fixed energy, rather the beam will have distribution of photons with energy varying over an allowed range (which depends on the initial electron and laser photon energies among other things). In such colliders, the cross-section and other observables should, therefore, be properly folded with a luminosity distribution function to get the measurable quantities, as is done in Eq.~\ref{eqn:lumfold}.

 At ILC high energy, high luminosity photon beam is obtained by Compton backscattering of low energy, high intensity laser beam off high energy electron beam. Ideal Compton backscattered photon spectrum is given by \cite{lumdist, photCollider,Muhlleitner:2005pr}
\begin{widetext}
\begin{eqnarray}
           f_{\gamma/e}(x) &=& \frac{1}{D(\xi)}~\left[1-x+\frac{1}{1-x}-4~\frac{x}{\xi (1-x)}+4~\frac{x^2}{\xi^2(1-x)^2}\right] \nonumber\\
           D(\xi)&=&\left(1-\frac{4}{\xi}-\frac{8}{\xi^2}\right)~\ln\left(1+\xi\right)
           +\frac{1}{2}+\frac{8}{\xi}-\frac{1}{2(1+\xi)^2},
\label{eqn:lumspec}
\end{eqnarray}
\end{widetext}
where $x=\frac{\omega}{E_e}$, with $E_e$ the energy of the initial electron and $\omega$  the energy of the scattered photon. $x$ thus gives the fraction of the electron energy carried by the scattered photon. 
Dependence of the distributions on the initial laser photon energy ($\omega_0$) 
comes through $\xi\approx \frac{4E_e\omega_0}{m_e^2}$, where $m_e$ is the electron mass. 
The maximum value of $x$ is $x_{max}=\frac{\xi}{1+\xi}$.  It is, but not possible to increase $\omega_0$ and $E_e $ to any value to get larger $x_{max}$. It is found that for $\xi$ beyond 
$\sim 4.8$, conversion efficiency drops down drastically due to $e^+e^-$ pair production between the laser photons and the backscattered photons, setting an absolute upper limit on $x\approx 0.83$. This value essentially means that with an electron beam of energy 
$E_e=250$ GeV, we can effectively go up to $\omega_0\approx 1.26$ eV . 

For an $e\gamma$ collider, the luminosity factor (Eq.~\ref{eqn:lumfold}) is the same as the above energy spectrum:
 \begin{equation}
\frac{d{\cal L}_{\gamma/e}(x)}{dx} = f_{\gamma/e}(x)
\label{eqn:lumfold}
\end{equation}

In a realistic collider, one need to also worry about many detailed aspects, like the non-linear effects making the actual photon spectrum deviating from the ideal case in Eq.~\ref{eqn:lumspec}, the polarization of the hard photon. The polarization ($P_\gamma$) of the scattered photon itself depend on the initial electron ($P_e$) and the laser beam polarization ($P_l$). In the high energy region, it is noted that $P_\gamma=-P_e$ \cite{lumdist, photCollider,Muhlleitner:2005pr}. This could be used to the advantage of the physics studies being considered. In our analysis, we have not gone into these details. we have assumed  unpolarized initial electron and laser beams, which produces an unpolarized high energy photon beam (to a very good approximation). 

For more details on photon collider one may refer to Ref.~\cite{lumdist, photCollider, Muhlleitner:2005pr, MoortgatPick:2005cw} and references therein.


\begin{thebibliography}{abc}
\bibitem{atlas}
  G.~Aad {\it et al.}  [ATLAS Collaboration],
  Phys.\ Lett.\ B {\bf 716}, 1 (2012)
  [arXiv:1207.7214 [hep-ex]].

\bibitem{Aad:2013wqa}
  G.~Aad {\it et al.}  [ATLAS Collaboration],
  Phys.\ Lett.\ B {\bf 726} (2013) 88
  [arXiv:1307.1427 [hep-ex]].
\bibitem{cms}
  S.~Chatrchyan {\it et al.}  [CMS Collaboration],
  Phys.\ Lett.\ B {\bf 716}, 30 (2012)
  [arXiv:1207.7235 [hep-ex]].

\bibitem{Chatrchyan:2013lba}
  S.~Chatrchyan {\it et al.}  [CMS Collaboration],
  JHEP {\bf 1306} (2013) 081
  [arXiv:1303.4571 [hep-ex]].

\bibitem{Dawson-EWSB}
  S. Dawson, Introduction to EWSB, (arXiv:hep-ph /9901280v1, 12 Jan 1999).

\bibitem{Djouadi:2005gi}
  A.~Djouadi,
  Phys.\ Rept.\  {\bf 457} (2008) 1
  [hep-ph/0503172].

\bibitem{Buchmuller:2006zu}
  W.~Buchmuller and C.~Ludeling,
  hep-ph/0609174.
\bibitem{Corbett:2013pja}
  T.~Corbett, O.~J.~P.~Éboli, J.~Gonzalez-Fraile and M.~C.~Gonzalez-Garcia,
  Phys.\ Rev.\ Lett.\  {\bf 111} (2013) 011801
  [arXiv:1304.1151 [hep-ph]].
\bibitem{Gonzalez-Fraile:2014cya}
  J.~Gonzalez-Fraile,
  arXiv:1411.5364 [hep-ph].
\bibitem{ILC}
  J.~Brau {\it et al.}  [ILC Collaboration],
  arXiv:0712.1950 [physics.acc-ph];
  G.~Aarons {\it et al.}  [ILC Collaboration],
  arXiv:0709.1893 [hep-ph].

\bibitem{Asner:2013psa}
  D.~M.~Asner, T.~Barklow, C.~Calancha, K.~Fujii, N.~Graf, H.~E.~Haber, A.~Ishikawa and S.~Kanemura {\it et al.},
  arXiv:1310.0763 [hep-ph].

\bibitem{couplings}
  K.~Moning, J.~Sekaric,
  Eur.\ Phys.\ J \ C {\bf 38}, 427-436 (2005).

\bibitem{pheno:egamWnu}
  S.~J.~Brodsky, T.~G.~Rizzo and I.~Schmidt,
  Phys.\ Rev.\  D {\bf 52}, 4929 (1995)
  [arXiv:hep-ph/9505441].  
  
\bibitem{pheno:Godfrey}
  M.~A.~Doncheski, S.~Godfrey and K.~A.~Peterson,
  arXiv:hep-ph/9710299;
  M.~A.~Doncheski, S.~Godfrey and K.~A.~Peterson,
  Phys.\ Rev.\  D {\bf 59}, 117301 (1999).  

\bibitem{pheno:Gregores}
  E.~M.~Gregores, M.~C.~Gonzalez-Garcia and S.~F.~Novaes,
  Phys.\ Rev.\  D {\bf 56}, 2920 (1997)
  [arXiv:hep-ph/9703430];

\bibitem{ww}
K.~Ackerstaff {\it et al.} [OPAL Collaboration], Eur.\ Phys.\ J.C2:597-606 (1998).

\bibitem{gam-gam2ww}
  E.~Yehudai Phys.\ Rev.\  D {\bf 44}, 3434 (1991).

\bibitem{Gounaris:1995mc}
  G.~J.~Gounaris, J.~Layssac and F.~M.~Renard,
  Z.\ Phys.\  C {\bf 69}, 505 (1996)
  [arXiv:hep-ph/9505430];

\bibitem{Hagiwara}
  K.~Hagiwara, R.~D.~Peccei, D.~Zeppenfeld and K.~Hikasa,
  Nucl.\ Phys.\  B {\bf 282}, 253 (1987).
  
  

\bibitem{LEPconstraints}
  J.~Alcaraz {\it et al.}  [ALEPH Collaboration and DELPHI Collaboration and
                  L3 Collaboration and OPAL Collaboration and LEP Electroweak Working Group ],
  arXiv:hep-ex/0612034.

\bibitem{Schael:2013ita} 
  S.~Schael {\it et al.}  [ALEPH and DELPHI and L3 and OPAL and LEP Electroweak Collaborations],
  Phys.\ Rept.\  {\bf 532}, 119 (2013)
  [arXiv:1302.3415 [hep-ex]].

\bibitem{CDFconstraints}
  V.~M.~Abazov {\it et al.}  [D0 Collaboration],
  Phys.\ Rev.\ Lett.\  {\bf 100}, 241805 (2008)
  [arXiv:0803.0030 [hep-ex]].

\bibitem{Chatrchyan:2011rr}
  S.~Chatrchyan {\it et al.}  [CMS Collaboration],
  Phys.\ Lett.\ B {\bf 701} (2011) 535
  [arXiv:1105.2758 [hep-ex]].

\bibitem{e-gamma collision}
  S.~Atag and I.~Sahin,
  Phys.\ Rev.\  D {\bf 75}, 073003 (2007)
  [arXiv:0703201 [hep-ph]].
  
\bibitem{LHCstudies}
  I.~Sahin and A.~A.~Billur,
  Phys.\ Rev.\  D {\bf 83}, 035011 (2011)
  [arXiv:1101.4998 [hep-ph]];
  E.~Chapon, C.~Royon and O.~Kepka,
  Phys.\ Rev.\  D {\bf 81}, 074003 (2010)
  [arXiv:0912.5161 [hep-ph]].

\bibitem{ILCstudies}
  E.~Yehudai,
  Phys.\ Rev.\  D {\bf 41}, 33 (1990);
  S.~Y.~Choi, J.~S.~Shim, H.~S.~Song, J.~Song and C.~Yu,
  Phys.\ Rev.\  D {\bf 60}, 013007 (1999)
  [arXiv:hep-ph/9901368]
  M.~Gintner, S.~Godfrey and G.~Couture,
  Phys.\ Rev.\  D {\bf 52} (1995) 6249
  [arXiv:hep-ph/9511204];
  H.~Aihara {\it et al.},
  arXiv:hep-ph/9503425;
  G.~Couture and S.~Godfrey,
  Phys.\ Rev.\  D {\bf 50}, 5607 (1994)
  [arXiv:hep-ph/9406257];
%
  Mamta Dahiya, Sukanta Dutta, Rashidul Islam,
  arXiv:1311.4523 [hep-ph].

  \bibitem{FORM}
  J.~A.~M.~Vermaseren,
  arXiv:math-ph/0010025.

\bibitem{CUBA} 
  T.~Hahn,
  Comput.\ Phys.\ Commun.\  {\bf 168}, 78 (2005)
  [hep-ph/0404043].
  
  \bibitem{lumdist}
  I.~F. Ginzburg {\it et al.}, Nucl. Instrum. Methods, {\bf 205}, 47 (1983); {\b 219}, 5 (1984).
  
  \bibitem{photCollider}
  V.~I.~Telnov,
  arXiv:0908.3136 [physics.acc-ph].

\bibitem{Muhlleitner:2005pr}
  M.~M.~Muhlleitner and P.~M.~Zerwas,
  Acta Phys.\ Polon.\  B {\bf 37}, 1021 (2006)
  [arXiv:hep-ph/0511339].


\bibitem{Beringer}
  J.~Beringer {\it et al.}, (Particle Data Group), 
  Phys.\ Rev.\  D {\bf 86}, 010001 (2012).

\bibitem{MoortgatPick:2005cw}
  G.~Moortgat-Pick, T.~Abe, G.~Alexander, B.~Ananthanarayan, A.~A.~Babich, V.~Bharadwaj, D.~Barber and A.~Bartl {\it et al.},
  Phys.\ Rept.\  {\bf 460} (2008) 131
  [hep-ph/0507011].

  \end{thebibliography}
\end{document}